\documentclass[%
 reprint,
 amsmath,amssymb,
 aps,
]{revtex4-2}

\usepackage{soul}

\usepackage{graphicx}
\usepackage{dcolumn}
\usepackage{bm}
\usepackage{xcolor}
\usepackage{multirow}

\usepackage{hyperref}
\usepackage[normalem]{ulem}
\hypersetup{
    colorlinks=true,
    linkcolor=blue,
    filecolor=magenta,      
    urlcolor=cyan,
    pdftitle={Overleaf Example},
    pdfpagemode=FullScreen,
    }

\begin{document}

\title{Adaptive Estimation of Drifting Noise in Quantum Error Correction}

\author{Devansh Bhardwaj}\email{ devansh.bhardwaj@duke.edu}
\affiliation{Duke Quantum Center, Duke University, Durham, NC 27701, USA}
\affiliation{Department of Electrical and Computer Engineering, Duke University, Durham, NC 27708, USA}

\author{Evangelia Takou}
\affiliation{Duke Quantum Center, Duke University, Durham, NC 27701, USA \mbox{}}
\affiliation{Department of Electrical and Computer Engineering, Duke University, Durham, NC 27708, USA}

\author{Yingjia Lin}
\affiliation{Duke Quantum Center, Duke University, Durham, NC 27701, USA}
\affiliation{Department of Physics, Duke University, Durham, NC 27708, USA}

\author{Kenneth R. Brown}\email{kenneth.r.brown@duke.edu}
\affiliation{Duke Quantum Center, Duke University, Durham, NC 27701, USA}
\affiliation{Department of Electrical and Computer Engineering, Duke University, Durham, NC 27708, USA}
\affiliation{Department of Physics, Duke University, Durham, NC 27708, USA}\affiliation{Department of Chemistry, Duke University, Durham, NC 27708, USA}

\date{\today}

\begin{abstract}
Advancing quantum information processors and building fault-tolerant architectures rely 
 on the ability to accurately characterize the noise sources  and suppress their impact on quantum devices. In practice, noise often drifts over time, whereas conventional noise characterization and decoding methods typically assume stationarity or provide only a time-average behavior of the noise. This treatment can result in suboptimal decoding performance. In this work, we present a rigorous analytical framework to capture time-dependent Pauli noise, by exploiting the syndrome statistics of quantum error correction experiments. 
We propose a sliding-window estimation method which allows us to recover the frequency components of the noise, by using optimal window sizes that we derive analytically. We prove the noise-filtering behavior of sliding windows, linking window size to spectral cutoff frequencies, and provide an iterative algorithm that captures multiple drift frequencies. We further introduce an overlapping window algorithm that enables us to capture rapid multi-frequency noise drifts in a single-pass fashion. Simulations for both phenomenological and circuit-level noise models validate our framework, demonstrating robust tracking of multi-frequency drift. The logical error rate obtained from our estimated models
consistently align with the ground-truth logical error rate, and we find suppression of logical errors compared to static error models.
 Our window-based estimation methods and adaptive decoding offer new insights into noise spectroscopy and decoder optimization under drift using only syndrome data.
\end{abstract}

\maketitle

\section{Introduction}
Building useful quantum computing architectures requires precise manipulation of quantum processors and protection of the  information they carry from unwanted error sources. Quantum error correction (QEC), deals with the noise by using multiple physical qubits to encode logical  information, making it less sensitive to the impact of the noise. Several recent works have demonstrated the success of QEC in achieving below threshold performance~\cite{google2025quantum,bluvstein2025architectural, reichardt2024demonstration}, or
realizing logical magic state distillation
\cite{sales2025experimental, daguerre2025experimental,dasu2025breaking}. QEC experiments demand accurate and fast decoders to decode the errors and yield reduced logical error rates. Optimal decoding performance can be achieved by feeding the decoder information about the noise profile of the device {\color{black}{such as drifts~\cite{Wang2023DGR}, noise bias~\cite{FlammiaPRX2019,FlammiaPRL2020,DuaPRXQ2023,campos2024clifford}, or device correlations~\cite{TakitaPRL2022,Remm2025}.}}

A major issue across all quantum computing platforms is that noise can drift over time\cite{kim2025error,Proctor2020,PhysRevB.111.115303,pelofske2023noise,10313742}. Time-varying fluctuations arise from naturally occurring experimental instabilities such as magnetic field and temperature variations \cite{mills2022impact,bermudez2017assessing}
, or spectral diffusion, which leads to fluctuating relaxation times~\cite{carroll2022dynamics}. Other sources of errors such as  calibration drift in control electronics, heating from cryogenic wiring, further introduce time-dependent gate errors~\cite{simbierowicz2024inherent}.
{\color{black}{These drifts manifest differently across frequency regimes. At low frequencies ($<1$ Hz), $1/f$ noise typically originates from TLS in superconducting qubits and from defect motion in spin qubits \cite{Paladino2014}. In the intermediate range (1 Hz–100 kHz), noise spectra often follow $1/f^2$ or Lorentzian forms, which arise from charge fluctuators and flux noise \cite{Koch2007,Krantz2019}. At high frequencies ($>1$ MHz), noise approaches a white spectrum, primarily driven by photon shot noise and Johnson–Nyquist processes \cite{Clerk2010}.}} While  noise drifts can be in part suppressed by regular calibration routines,
those cannot be run in parallel with QEC experiments, making noise drift an issue for the logical error performance. Hence, if drifting noise is not accounted for, the decoding performance degrades.

Certain noise characterization methods can identify noise drifts, for example quantum process tomography \cite{tinkey2021quantum,Proctor2020} or randomized benchmarking \cite{qi2021randomized,gaye2025model}. However, these experiments increase the experimental and computational overhead in their effort to provide a good model for decoding QEC experiments. A more effective solution is to estimate directly the noise based on syndromes. While several works have pursued this direction assuming static noise ~\cite{Remm2025,takou2025estimating,BlumeKohout2025}, it still remains an open question whether drifts in the error rates relevant to decoders can be captured in a similar manner using only the syndrome history.

In this work, we address the challenge of learning time-dependent error rates using only the syndrome history of QEC experiments. Unlike the approach in Ref~\cite{Spitz2018}, our framework can capture an arbitrary number of frequency components in the noise spectrum. Our methods can be applied  to any quantum error-correcting code or physical Pauli error model, and allow us to create directly a noise-aware detector error model, compressing thus the information that needs to be provided to the decoder. 
The first technique we develop is based on sliding windows that we apply on the syndrome data to capture frequency components below a certain frequency cutoff. We prove that this method acts as a low-pass filter with frequency cutoff determined by the window size. By iterating the sliding window process with varying window sizes, we can capture multiple frequency components. We further supplement this method with analytical results regarding optimal window sizes
for accurate estimation, and the expected statistical uncertainties in the error rates.  Our analytical results can further explain time-lag decoding issues which have been observed previously in Ref~\cite{Spitz2018}. Additionally, we develop a relative window estimation method which allows us to capture more rapidly drifting error rates. This technique is a single-pass procedure, meaning we do not need to iterate it multiple times to get all relevant frequency components. Through these methods, we successfully estimate a wide variety of noise drift profiles and demonstrate that the resulting logical error rate is consistent with logical error rate obtained from the ground truth model. Additionally, we find that our estimated detector error models suppress logical errors better than approximate static detector error models. Our approach is general and captures time-dependent noise without extra experimental cost and without relying on machine learning or heuristic methods.

The paper is structured as follows. In Sec. \ref{Overview of noise estimation based on syndrome history}, we review noise estimation based on syndrome history for static error rates. In Sec. \ref{Capturing time-dependence of error rates}, we introduce our sliding-window and relative-window estimation methods which capture time-dependent error rates.
 In Sec.  \ref{Results}, we show our simulations results for various noise-drift scenarios.  In Sec. \ref{Decoding Results}, we study how noise-drifts impacts the decoding performance. In Sec.~\ref{Discussion}, we discuss the limitations inherent in existing approaches, followed by potential extensions of our framework and avenues for future investigation. Finally, in Sec.~\ref{Conclusion}  we conclude our results.

\section{Overview of noise estimation based on syndrome history}\label{Overview of noise estimation based on syndrome history}

In QEC experiments, noise models can be described by detector error models (DEM) based on detectors that track parity changes in repeated measurements of the same stabilizer across adjacent time steps~\cite{derks2024designing,Gidney2021stimfaststabilizer}. Graphically, a DEM can be represented by a decoding graph (or hypergraph). In this work, we consider only decoding graphs. A decoding graph is a weighted graph with detectors as nodes, and error mechanisms as edges. If two detectors are connected by an edge, the detectors are flipped if the error mechanism corresponds to the edge is triggered. These edges connecting two detectors are called bulk edges. Additionally, we introduce an artificial boundary node. If an edge connects a detector and the boundary node, the corresponding error mechanism can only flip this particular detector. Such edges are called boundary edges. The weight associated to each edge is given by $w=\ln((1-p)/p)$ where $p$ is the probability that this error mechanism occurs. 
Alternatively, a DEM can be represented using a list of independent error events that flip subsets of detectors, together with the probabilities that these errors happen. Typically, the DEM is generated by a circuit error model ( qubit or gate error model) for example via Stim~\cite{Gidney2021stimfaststabilizer}. By abstracting away the circuit-level details of the noise channel, the DEM supplies the statistical information needed for the decoder to infer the most likely error based on the observed syndrome history. Therefore, an optimal decoder and error correction strategy requires knowledge of the DEM. 

Recently, the problem of learning the error rates of a DEM using the syndrome history of QEC experiments~\cite{BlumeKohout2025,takou2025estimating,Remm2025,Spitz2018} has received significant attention as a noise characterization method that avoids the need of extra benchmarking or characterization experiments. At the same time, knowing the DEM can enhance decoding performance via noise-aware decoding~\cite{takou2025estimating}. 

To learn the DEM probabilities, syndromes
from several syndrome extraction cycles and repetitions of the same experiment are collected. Based on the syndromes, one can count the average number of times a detector fires, coincidences of detectors, or even higher-order correlators (which are relevant only for decoding hypergraphs), and use them to estimate the error rates. 

Ref.~\cite{Spitz2018} showed that {\color{black}{in the case of independent event probabilities (which holds for DEM models undergoing Pauli noise),}} the probability $p_{ij}$ of a bulk error ( a bulk edge incident to detectors $v_i$ and $v_j$ in the graph) is given by:

\begin{equation}
p_{ij} = \frac{1}{2} - \sqrt{\frac{1}{4} - \frac{\langle{v_i v_j}\rangle - \langle{v_i}\rangle\langle{v_j}\rangle}{1 - 2(\langle{v_i}\rangle + \langle{v_j}\rangle) + 4\langle{v_i v_j}\rangle}},
\label{eq:bulk_edge}
\end{equation}
whereas the probability $p_{ii}$ of a boundary edge (an error that flips a boundary detector $v_i$)
is given by:

\begin{equation}
p_{ii} = \frac{1}{2} + \frac{\langle v_i \rangle - 1/2}{\prod_{j \ne i} (1 - 2p_{ij})}.
\label{eq:bound_edge}
\end{equation}
Here, $\langle \cdot \rangle$ denotes the expectation value: $\langle v_i\rangle$ counts the number of times a single detector fires, and $\langle v_iv_j\rangle$ counts how often two detectors fire together, each normalized by the total number of experimental shots ( number of QEC experiments). 
Thus, by using the entire syndrome history from multiple experimental trials, one can obtain the error rates of the DEM.

{\color{black}{
When the above equations are evaluated using the full syndrome history, only time-average error rates can be obtained. 
As a result, the time-dependent noise profile is effectively lost. 
Assuming that the error rates remain approximately constant across syndrome extraction cycles or experiments might not always be a valid assumption in practice, and might lead to suboptimal decoding performance.
Thus, capturing the time-dependent profile could provide insights into the device's performance and its connection to decoding performance.
}}

So far, previous works that studied the estimation of Pauli error rates from syndromes~\cite{Spitz2018, takou2025estimating,BlumeKohout2025} did not consider temporal variations for the error rates. {\color{black}{In other words, the full syndrome history was used to obtain the estimated probabilities based on Eq.~\eqref{eq:bulk_edge} and Eq.~\eqref{eq:bound_edge}.
In what follows next, we show how to estimate the time-dependence of error rates, via new post-processing methods we propose, but still using the syndrome data like previous works.
}}

\section{Capturing time-dependence of error rates} \label{Capturing time-dependence of error rates}
{\color{black}{
In this section, we will describe our three methods to estimate detector error models with time-dependent error rates. All these methods are based on the same principle of using windows that capture part of the syndrome history. By processing detection events within specified windows and not across the full syndrome history, we are able to retain the time-dependent signal. The three methods which we will analyze shortly below are summarized in Fig.~\ref{fig:st_graph}}}.

\subsection{Sliding Window Estimation} \label{sub:Sliding Window Estimation}

\begin{figure*}[!ht]
    \centering
    \includegraphics[width=1.01\linewidth]{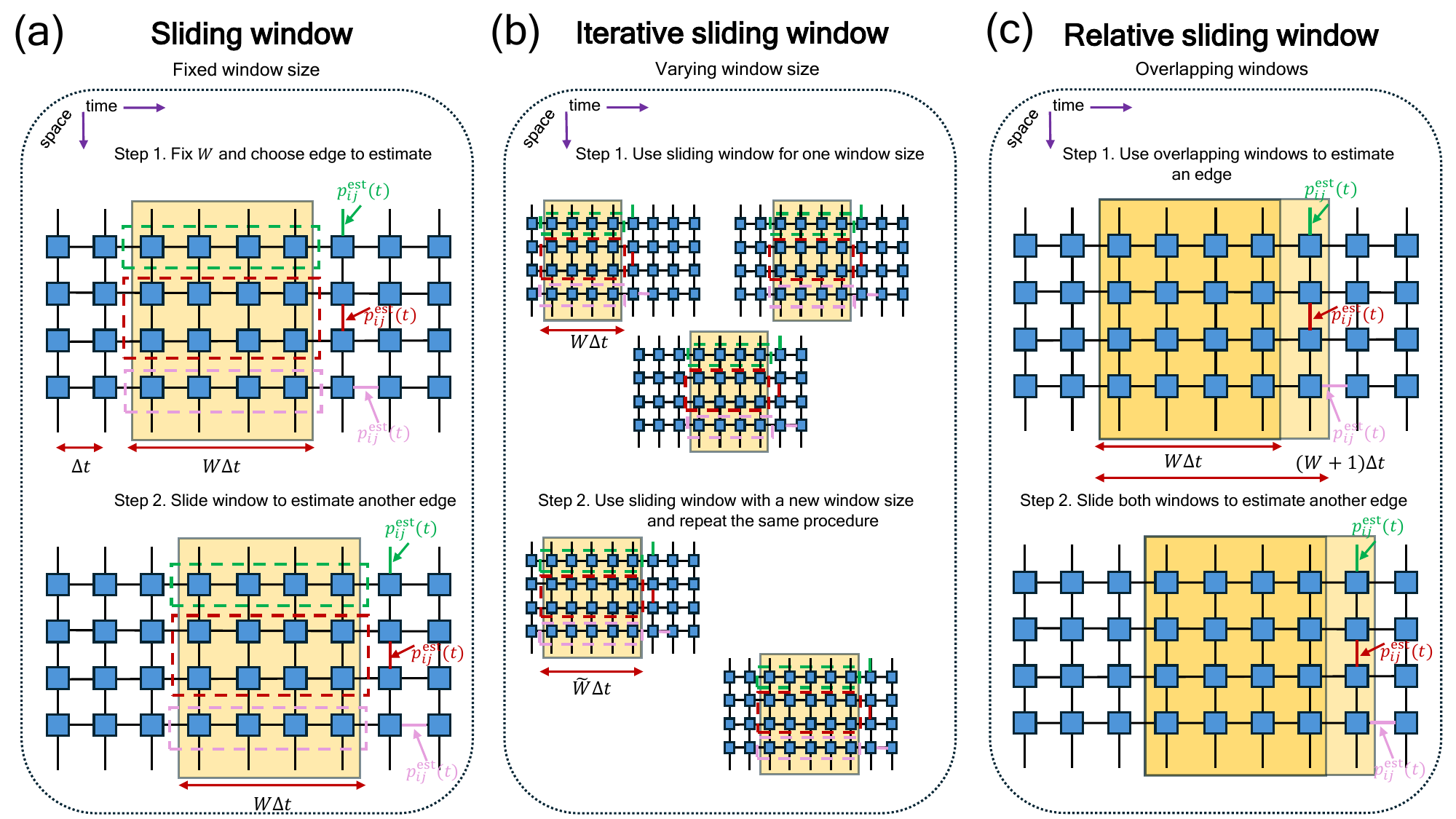}
    \caption{Schematic overview of our methods for estimating time-varying noise from syndrome data of memory QEC experiments. (a) Sliding window estimation: A window of fixed size slides along the temporal axis of the syndrome data. Time-varying error rates, $p_{ij}^{\text{est}}(t)$, are estimated from the syndromes within the window at each position. (b) Iterative sliding window estimation: The sliding window estimation  process iterates over a range of window sizes. Starting with an initial window, the method sequentially uses the output of one iteration as a prior for the next, refining the estimate with each step. This method aggregates information across multiple windows to estimate the error rate signal $p_{ij}^{\text{est}}(t)$. (c) Relative window estimation: Two overlapping windows of sizes $W$ and $W+1$ slide simultaneously along the temporal axis. The differential information between the two windows is analyzed to estimate the time-varying error rates.}
    \label{fig:st_graph}
\end{figure*}

We begin with the first method we developed that we term sliding window estimation. The key feature of this approach is that we use expectation values of detection events within a fixed time window of $W$ syndrome extraction cycles, namely in some time interval $[t_l - W\Delta t, t_l)$, as shown in Fig.~\ref{fig:st_graph}. Here, the time index $t_l$ refers to the time instance where the syndrome has been obtained from the $l$-th syndrome extraction cycle. Given the time interval between two syndrome extraction cycles is $\Delta t$, we label the syndrome extraction times as $t_n = n\Delta t$, with $n\in[0,N-1]$ denoting the syndrome extraction cycle number .

Applying Eq.~(\ref{eq:bulk_edge}) and Eq.~(\ref{eq:bound_edge}) for a time window $[t_l - W\Delta t, t_l)$ gives us the estimated probability at $t_l$, $p_{ij}^{\text{est}}(t_l)$. Thus, by choosing a different  time $t_l$, and sliding the window of the same size $W$, we can estimate the error rates at various points in time and space.
In Appendix \ref{Appendix A}, we prove that the estimated probability relates to the true probabilities via the expression:

\begin{equation}
p_{ij}^{\mathrm{est}}(t_l) = \frac{1}{W} \sum_{k=0}^{W-1} p_{ij}(t_l - W\Delta t + k\Delta t),
\label{eq:temp_average}
\end{equation}
where $k$ indexes the syndrome extraction cycles within
the window. This results shows that $p_{ij}^{\mathrm{est}}(t_l)$
is the temporal average of ground-truth event probabilities $p_{ij}(t)$ {\color{black}{within a specified window}}. 
As per the structure of our proof, the relation in Eq.~\eqref{eq:temp_average} holds for Markovian time-dependent noise.

Since the sliding window estimation is based on statistics of detection events, the estimated probabilities have an associated statistical uncertainty. In Appendix \ref{Appendix B}, we show that the standard deviation of $p_{ij}^{\mathrm{est}}(t_l)$ within a window of size $W$ is given by:

\begin{equation}
\begin{split}
\sigma_W  &= \frac{1}{W} \times 
\\&\sqrt{\sum_{k=0}^{W-1} p_{ij}(
(l-W+k)\Delta t)
\left(1 - p_{ij}((l-W+k)\Delta t)\right)}.
\label{eq:Stand_deviation}
\end{split}
\end{equation}
This expression accounts for the binomial variance of each sample in the window, scaled by the squared window size due to averaging.

\begin{figure*}[!htbp]
    \centering
    \includegraphics[width=1.02\linewidth]{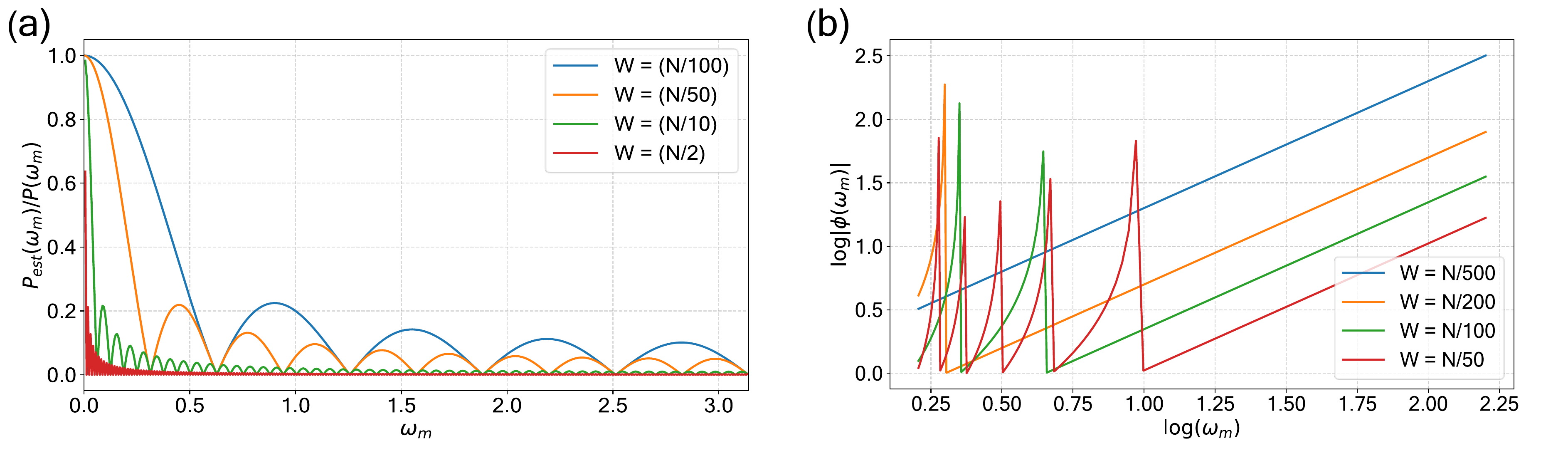}
    \caption{ Impact of window size on noise spectrum estimation. (a) Ratio of the estimated noise spectrum, $P_{\text{est}}(\omega_m)$, to the true noise spectrum, $P(\omega_m)$, across different frequencies, demonstrating the damping introduced by various window sizes $W$. (b) Log-log plot of the phase spectrum, $\phi(\omega_m)$, against frequency, $\omega_m$, for different window sizes. Window sizes are defined as fractions of the total syndrome extraction cycles, $N$.}
    \label{fig:tf_response}
\end{figure*}

We can gain further insight of how the sliding window estimation method behaves if we turn our attention to the frequency-domain representation of the error rates.  Applying the discrete Fourier transform (DFT) to Eq.~(\ref{eq:temp_average}) we find:
\begin{equation}
p_{ij}^{\mathrm{est}}(\omega_m) = \frac{1}{W} \sum_{n=0}^{N-1} \sum_{k=0}^{W-1} p_{ij}(n - W + k) e^{-i2\pi  m n/N}.
\end{equation}
Here, $\omega_m = 2\pi m/(N\Delta t)$ (for $m = 0,...,N-1$) denotes the discrete angular frequency, $N$ is the total number of syndrome extraction cycles, and $W$ is the window size. This expression captures the spectral characteristics of the probability estimator through its dependence on the sliding window. The frequency-domain representation of the estimator further simplifies to:
\begin{equation}
p_{ij}^{\mathrm{est}}(\omega_m) = \frac{1}{W} \left(\frac{1 - e^{-i2\pi  m W/N}}{1 - e^{-i2\pi  m/N}}\right) p_{ij}(\omega_m),
\end{equation}
 This can be expressed in standard magnitude-phase form as:
\begin{equation}\label{Eq:pijOmega}
p_{ij}^{\mathrm{est}}(\omega_m) = {\frac{1}{W} \left|\frac{\sin(\pi m W/N)}{\sin(\pi m/N)}\right|} \cdot {e^{i\pi m(W-1)/N}} \cdot p_{ij}(\omega_m)
\end{equation}
revealing the spectral modulation introduced by the sliding-window averaging process. The magnitude term in Eq.~(\ref{Eq:pijOmega}) represents the window's frequency response, while the phase factor accounts for the temporal shift.

The frequency analysis in Eq.~\eqref{Eq:pijOmega} reveals how the  accuracy 
of noise estimation depends on window size $W$. The magnitude response of the estimated Fourier spectrum follows the standard Dirichlet kernel form which exhibits characteristic sidelobe attenuation at higher frequencies ($\omega_m = 2\pi m/N\Delta t$ for $m \gg 1$). This {means that our sliding-window estimation on syndrome data acts as a discrete low-pass filter, suppressing high-frequency noise in the estimated probabilities. The filter introduces a  phase shift, $\phi= \frac{\pi m(W-1)}{N}$,
as derived in Eq.~\eqref{Eq:pijOmega}. For a given window size $W$,  the time-lag introduced by the estimator is periodic with respect to the noise spectrum frequencies, $\omega_m$. 
Figure \ref{fig:tf_response} illustrates both the magnitude roll-off and phase characteristics based on Eq.~(\ref{Eq:pijOmega}), demonstrating the trade-off between estimation error suppression (governed by $W$) and temporal resolution.

Based on this frequency analysis, we can also identify
the optimal window size for frequency-selective estimation. Given a target cutoff frequency $\omega_c = 2\pi m_c/(N\Delta t)$ and maximum allowable estimation error $\epsilon$, the optimal window size $W_{\text{opt}}$ satisfies the  constraint:

\begin{equation}
\left|\frac{\sin(\pi m_c W_{\text{opt}}/N)}{\sin(\pi m_c/N)}\right| = 1 - \epsilon.
\label{eq:transf_fun}
\end{equation}
Using numerical methods we can further simplify this equation to the approximate solution:
\begin{equation}
W_{\text{opt}} \approx \frac{c(\epsilon)N}{m_c},
\label{eq:Optimal_window}
\end{equation}
where $c(\epsilon) $ is an $\epsilon$-dependent prefactor. The inverse relationship between $W_{\text{opt}}$ and $m_c$ shows the trade-off between frequency resolution and estimation accuracy.
The prefactor $c(\epsilon)$ in Eq.~(\ref{eq:Optimal_window}) is a positive constant bounded by $0 < c(\epsilon) \leq m_c$, which encodes the error tolerance $\epsilon$ in frequency estimation. For $\epsilon = 0.05$, numerical solution of the transcendental equation yields $c(0.05) \approx 0.12$.

By carefully selecting the window size \( W \) using the derived formulae, one can balance out the trade-off in accuracy and frequency resolution. From Eq.~\eqref{eq:Optimal_window}, we further note that the standard deviation in estimating \( p_{ij}(t) \) depends on both the error rates and the window size \( W \). 
Minimizing this deviation thus requires sufficiently large windows, but we cannot use extremely large windows since this would restrict our estimation to very slowly-drifting noise. Nevertheless, as we will verify later with our simulations, our sliding-window estimation can still capture reasonably fast noise drifts. Existing studies~\cite{Proctor2020} further suggest that noise drift predominantly comprises low-frequency components (\( N/m > 10^3 \)) that our method can easily handle.

Although the sliding-window estimation can properly capture single frequency  noise drift, an additional complication arises from the phase shift it induces, as quantified in Eq.~(\ref{Eq:pijOmega}). This phase shift implies that merely optimizing \( W \) is insufficient when the noise spectrum contains multiple frequency components, because it can create a lag in the estimated time signal. For noise drift dominated by a single frequency, a viable strategy is to employ a large window for estimation and subsequently apply the phase and magnitude corrections from Eq.~(\ref{Eq:pijOmega})  to recover the ground-truth signal. This approach not only mitigates estimation variance but also resolves lag features. Our analytical expressions also explain the  time lag observed for large window sizes in Ref.~\cite{Spitz2018}, aligning our theoretical proof with empirical results.
{\color{black}{While the above correction strategy can resolve time lag features when a single frequency dominates in the noise spectrum, this is not the case for multi-frequency drifts. In the latter case,}}
the direct application of phase and magnitude corrections becomes inaccurate, since these corrections are inherently frequency-dependent. To address this limitation, we introduce next an iterative sliding-window approach.
\subsection{Iterative sliding window estimation \label{Sec:IterativeSliding}}
To estimate a DEM whose error rates drift with multiple distinct frequencies, we propose an iterative sliding-window approach, illustrated in Fig.~\ref{fig:st_graph}(b), that leverages the inherent low-pass filtering behavior of a sliding-window. {\color{black}{The main idea of this method is that we}} initially use a large window $W_0 \sim \mathcal{O}(N)$, which is designed to capture low-frequency components, and iteratively reduce the window size ($W_k > W_{k+1}$) to resolve higher-frequency bands. {\color{black}{Here $k$ labels the iterative window sizes we pick, with $k=0$ being the largest window size.}} {\color{black}{Each window size  captures frequencies below a cutoff, but can still have small contributions from higher frequency components. These contributions, however, will be negligible for larger windows, and can be further captured by smaller window sizes. To control this feature, we set a threshold of $\mu=1-\epsilon$, where we typically select $\mu\in[0.05,0.2]$, to effectively discard frequency components whose magnitude is below this threshold. }} 

{\color{black}{
In more detail, we initially start with a large window size $W_0$ and estimate the error rates via Eq.~\eqref{eq:bulk_edge} and Eq.~\eqref{eq:bound_edge}. By sliding this window size across the syndrome history, we then obtain the estimated $p_{ij}^{\text{est}}(t_l)$ values for various times $t_l$, which obey  Eq.~\eqref{eq:temp_average}. Based on the window size $W_0$ and the threshold $\mu$ we set, we obtain a cutoff frequency, $\omega_{c}^{(k=0)}$, based on Eq.~\eqref{eq:transf_fun} by rounding up that equation. This means, that we can then write the time-dependent probabilities we estimate for the particular window as:

\begin{equation}
\begin{split}
    &p_{ij}^{\text{est}}(t_l) = \frac{1}{W}  \sum_{m=0}^{m_{c}^{(k=0)}} \left( \frac{\sin(\omega_m W / 2)}{\sin(\omega_m / 2)} \right)\\
           &  \Bigg[ p_{a_m} \sin\left( (2t_l - (W - 1)\Delta t)\frac{\omega_m}{2} \right)  + \\ &p_{b_m} \cos\left( (2t_l - (W - 1)\Delta t)\frac{\omega_m}{2} \right) \Bigg].
\end{split}
\end{equation}

The reason why we can truncate the above formula to $m_{c}^{(k=0)}$ is because the contribution of higher frequencies will be negligible for this window size. Given the fact that we know the estimated values across various time steps, and the fact that the frequencies on the right hand side are discrete, we can apply the least squares method to obtain the solutions for the ground-truth probabilities $p_{a,m}$ and $p_{b,m}$. After finishing this procedure, we utilize a new window $W_1$, such that $W_0>W_1$. Once again, we determine the frequency cutoff, $\omega_{c}^{(k=1)}$, for the new window size; because of the low-pass action of the sliding-window method, it holds that $\omega_{c}^{(k+1)}\geq \omega_{c}^{(k)}$.
If after rounding up, the new frequency equals the previous cutoff frequency, then we repeat the estimation procedure we mentioned above and take the average of the $p_{a_m}$ and $p_{b_m}$ solutions from both windows. If the new cutoff frequency is larger than the previous one, then we we feed the previous $p_{a_m}$'s and $p_{b_m}$'s  we know for frequencies up to the previous cutoff, and then solve for the new $p_{a_m}$'s and $p_{b_m}$'s that remain unknowns for the smaller window size of this step. This procedure is iterated till we reach some final minimal window size, below which we cannot proceed due to very high statistical uncertainty in the error rates. This size is typically $500-1000$ syndrome extraction cycles.  

This iterative strategy balances spectral resolution and estimation precision and allows us to discriminate multiple frequencies. Due to the fact that we relate the estimated probabilities to the temporal averages, we additionally eliminate time-lag issues. These features will become apparent later on when we present our simulation results.

}}

\subsection{Relative Window Estimation}\label{Relative Window Estimation}

While iterative sliding window estimation provides fine resolution of individual frequency components, yielding valuable insight into device noise sources, it comes at the cost of multiple passes  and is limited to noise that is not drifting too rapidly ($N/m_c>10^3$).  {\color{black}{In situations where resolution of individual frequency components is less critical, we}} present an alternative method to estimate the drifting error rates, that we term relative window estimation. This method allows us to estimate  {\color{black}{even}} rapidly drifting error rates, without having to optimize for window sizes. In other words, it is a single-pass method through which we can capture the instantaneous error rates, which could comprise of multiple frequency components.   To estimate \( p_{ij}(t_l) \) at time \( t_l \), we analyze two overlapping windows:
\begin{itemize}
    \item Window of size \( W \): Covers \( W \) cycles in time interval \( t \in [t_l - W\Delta t, t_l) \)
    \item Window of size \( W+1 \): Covers \( W+1 \) cycles in time interval \( t \in [t_l - W\Delta t, t_l + \Delta t) \)
\end{itemize}

Based on Eq.~\eqref{eq:temp_average}, by selecting a window of size \( W \) that we use to obtain $p_{ij}^{\text{est}}(t)$, we capture the temporal average of \( p_{ij}(t) \). Crucially, the relative information captured by the larger window \( W+1 \) compared to \( W \) isolates the instantaneous probability at \( t_l \). This is because the additional time step in \( W+1 \) introduces only \( p_{ij}(t_l) \) as new information, while the overlapping interval \( [t_l - W\Delta t, t_l) \) contributes identically to both windows.  

Thus, the instantaneous probability is derived as:
\begin{equation}
p_{ij}(t_l) = (W+1) p_{ij, W+1}^{\mathrm{est}}(t_l + \Delta t) - W p_{ij,W}^{\mathrm{est}}(t_l),
\label{eq:relative window estimation}
\end{equation}
where $p_{ij,W}^{\mathrm{est}}(t_l) $ and $p_{ij,W+1}^{\mathrm{est}}(t_l+\Delta t) $ are the estimates from each window. Figure \ref{fig:st_graph}(c) illustrates this mechanism.
To estimate the instantaneous $p_{ij}(t)$ at arbitrary times, both overlapping windows of size $W$ and $W+1$ slide by a discrete time step $\Delta t$. {\color{black}{In what follows next, we will explore via numerical simulations the accuracy of the noise estimation methods we developed, as well as the decoding performance we obtain from the estimated DEMs.}}

\section{Results} \label{Results}
In this section we perform numerical simulations to test our estimation methods. All our simulations were performed using Stim~\cite{Gidney2021stimfaststabilizer}. For simplicity, we will be considering a repetition code memory and surface code memory, although our methods can be straightforwardly applied to other codes as well.

\begin{figure*}[!htbp]
    \centering
    \includegraphics[width=1.02\linewidth]{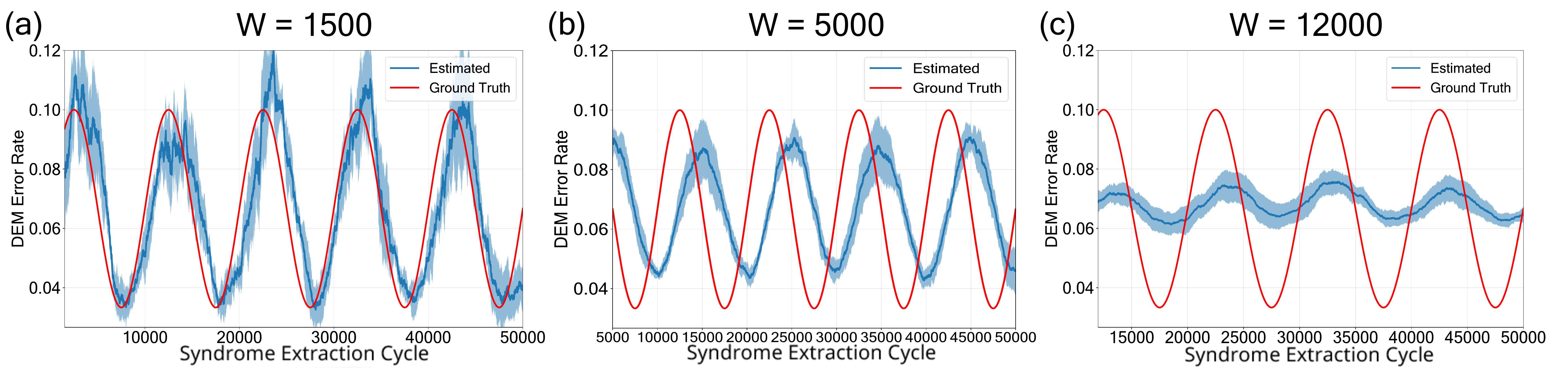}
    \caption{
    {\color{black}{Estimating DEM error rates using the sliding window approach, for a distance-3 repetition code under phenomenological noise model. Data and ancilla qubits experience depolarizing noise per syndrome extraction cycle with parameters $g_0=0.1$, $g_1=0.05$, and frequency $\omega_1 = 2\pi/(10^4 \Delta t$).
    (a) Estimated DEM error rate of bulk edge averaged over all qubits, as a function of the syndrome extraction cycles, for a window size of $W=1500$. The red curve shows the ground truth error rate, and the blue curve the estimated error rate. (b) Same as in (a) for a window size $W=5000$. (c) Same as in (a) for a window size $W=12000$. For the estimated signal, we repeat the simulation five times to plot the average behavior and display the standard deviation with the error bars.
    }}}
    \label{fig:filter_response}
\end{figure*}

\subsection{Application of sliding window estimation to a phenomenological noise model} \label{Filtering Action}
We begin by investigating the low-pass filtering behavior inherent to sliding window estimation. To demonstrate this, we consider a distance-3 repetition code under a phenomenological noise model. We assign time-varying single-qubit depolarizing errors at the start of each syndrome extraction cycle to all data and ancilla qubits, with the noise channel given by:
\begin{equation}
\mathcal{E}(\rho) = (1-g(t))\rho + \frac{g(t)}{3}(\sigma_x\rho\sigma_x + \sigma_y\rho\sigma_y + \sigma_z\rho\sigma_z),
\label{eq: 1q dep channel}
\end{equation}
where $\sigma_x$, $\sigma_y$, $\sigma_z$ are the Pauli matrices. We further define $g(t)$ as:
\begin{equation}
g(t) = g_0 + \sum_{m \in \mathcal{M}} g_m \sin(\omega_m t), ~g_0,g_m\in[0,1].
\label{eq: time_parameters}
\end{equation}
Here, $\mathcal{M}$ denotes a set of discrete frequencies. In the limit of $\omega_m=0, \forall m$, we recover a time-independent error rate given by $g_0$. The continuous-time probability $g(t)$ is discretized by taking the time difference between consecutive syndrome extraction cycle to be $\Delta t$, such that $t_n = n\Delta t$, where $n$ is the syndrome extraction cycle index.

We consider $5\times10^4$ syndrome extraction cycles where a single drift frequency is considered for all qubits, with parameters $g_0=0.1$, $g_1=0.05$, and frequency $\omega_1 = 2\pi/(10^4 \Delta t$). In Fig.~\ref{fig:filter_response},  estimated and ground-truth DEM error rate of the bulk edge is shown as a function of syndrome extraction cycles. The ground-truth error rate is shown by the red line, while the blue line represents the average estimated error rate obtained via the sliding-window approach. In Fig.~\ref{fig:filter_response}(a) we consider a window size of $W=1500$, in Fig.~\ref{fig:filter_response}(b) a size of $W=5000$, and in Fig.~\ref{fig:filter_response}(c) a size of $W=12000$. {\color{black}{In each case, we repeat the estimation experiment five times using the same parameters, in order to get the average behavior and show the error bars (standard deviation). }}{\color{black}{For the smallest window size, $W=1500$, we see that we capture the frequency of the signal with good accuracy. The estimated error rate's component, $g_1$, corresponding to frequency $\omega_1$, is damped by a factor of ${0.964=\frac{1}{1500}\left|\frac{\sin(3\pi/20)}{\sin(\pi/(10^4)}\right|}$ and is phase shifted by $3\pi /20 \approx\frac{\pi (1499)}{10^4}$ }}. The estimated error rate's spread, especially at the maxima,
is pronounced. This behavior can be understood based on Eq.~\eqref{eq:Stand_deviation}. The term $p_{ij}(t)(1-p_{ij}(t))$ in Eq.~\eqref{eq:Stand_deviation} is a concave function that achieves its maximum at $p_{ij}(t) = 0.5$. Given the assumption that error rates are below $0.5$, this term is monotonically increasing in this regime. Thus, the standard deviation of the estimate is higher for local maxima of the error rate function $p_{ij}(t)$. However, since the standard deviation is also inversely proportional to $\sqrt{W}$, a larger window size $W$ mitigates this effect, capturing the error rate over the entire interval with approximately uniform deviation. {\color{black}{As we increase the window size to $W=5000$, we observe that the spread reduces, and we again estimate the correct frequency, but the component, $g_1$, is damped by a factor of ${0.636=\frac{1}{5000} \left|\frac{\sin(\pi/2)}{\sin(\pi/(10^4)}\right|}$ and phase shifted by $\pi/2 \approx\frac{\pi (4999)}{10^4}$. }} Finally, for the largest window size, $W=12000$, as expected, we almost obtain the mean value of the sinusoidal signal 
accompanied by weak oscillations. {\color{black}{Precisely, the component, $g_1$, is damped by a factor of ${0.156=\frac{1}{12000} \left|\frac{\sin(\pi/5)}{\sin(\pi/(10^4)}\right|}$ and phase shifted by $\pi/5 \approx\frac{\pi (1999)}{10^4}$.}}  These results verify the mathematical framework of sliding window estimation we introduced in the previous sections in terms of frequency response, time-lag features and statistical uncertainties in the estimated error rates. 

\begin{figure*}[!htbp]
    \centering
    \includegraphics[width=1.02\linewidth]{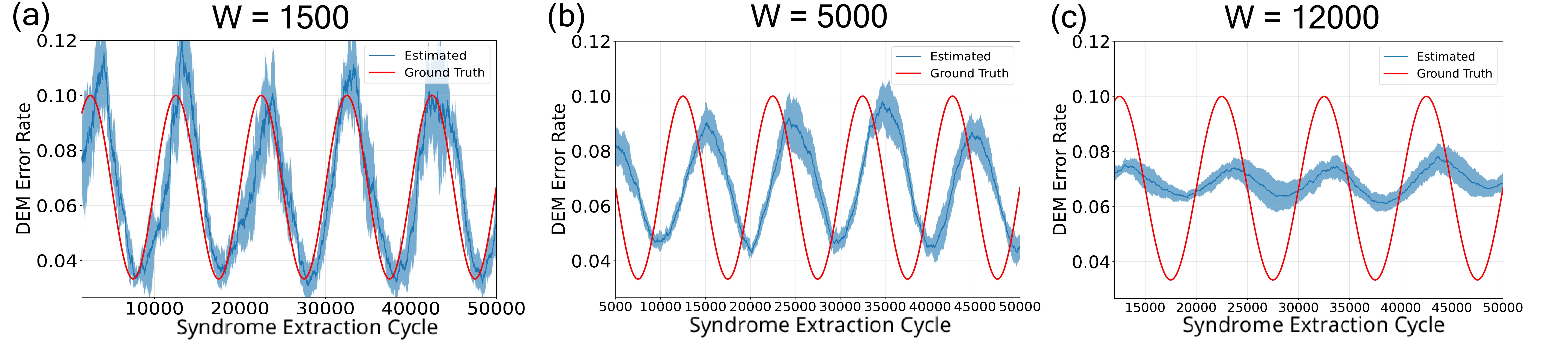}
    \caption{
    {\color{black}{Estimating  X-DEM error rates using the sliding window approach, for a distance-3 rotated surface code under phenomenological noise model. Data and ancilla qubits experience depolarizing noise per syndrome extraction cycle with parameters $g_0=0.1$, $g_1=0.05$, and and frequency $\omega_1 = 2\pi/(10^4 \Delta t$).
    (a) Estimated X-DEM error rate of bulk edge averaged over all qubits, as a function of syndrome extraction cycle, for a window size of $W=1500$. The red curve shows the ground truth error rate, and the blue curve the estimated error rate. (b) Same as in (a) for a window size $W=5000$. (c) Same as in (a) for a window size $W=12000$. For the estimated signal, we repeat the simulation five times to plot the average behavior and display the standard deviation with the error bars.
    }}}
    \label{fig:filter_response_sc}
\end{figure*}

{\color{black}{Next, we analyze a rotated surface code with code distance $d=3$ under a phenomenological noise model. The time-varying single-qubit depolarizing channel  is defined analogously to Eq.~\eqref{eq: 1q dep channel} and Eq.~\eqref{eq: time_parameters}} . We measure only the X-stabilizers and estimate the corresponding X-DEM, since for a surface code under the phenomenological noise model, the X-DEM and Z-DEM are equivalent. We simulate $5\times10^4$ syndrome extraction cycles, assuming a single drift frequency that is common to all data and ancilla qubits. The noise parameters are set to $g_0=0.1$, $g_1=0.05$, and frequency $\omega_1 = 2\pi/(10^4 \Delta t$).

Figure \ref{fig:filter_response_sc} presents the estimated and ground-truth X-DEM error rate of the bulk edge as functions of the syndrome extraction cycle. The ground-truth error rate is shown in red, while the average estimated error rate, obtained using the sliding-window approach, is shown in blue. To account for statistical variation, each simulation is repeated five times, and the resulting error bars are plotted accordingly.
Figure \ref{fig:filter_response_sc}(a), \ref{fig:filter_response_sc}(b) and \ref{fig:filter_response_sc}(c) display the estimation results for sliding-window sizes of 1500, 5000, and 12000, respectively. As anticipated, the sliding-window estimation method captures the ground-truth X-DEM error rate with the $g_1$ component exhibiting a similar damping and phase shift factors as observed for the repetition code. This consistency demonstrates the general applicability of the sliding-window estimation framework across different quantum error-correcting codes.
}}

\begin{figure*}[!ht]
    \centering
    \includegraphics[width=1.02\linewidth]{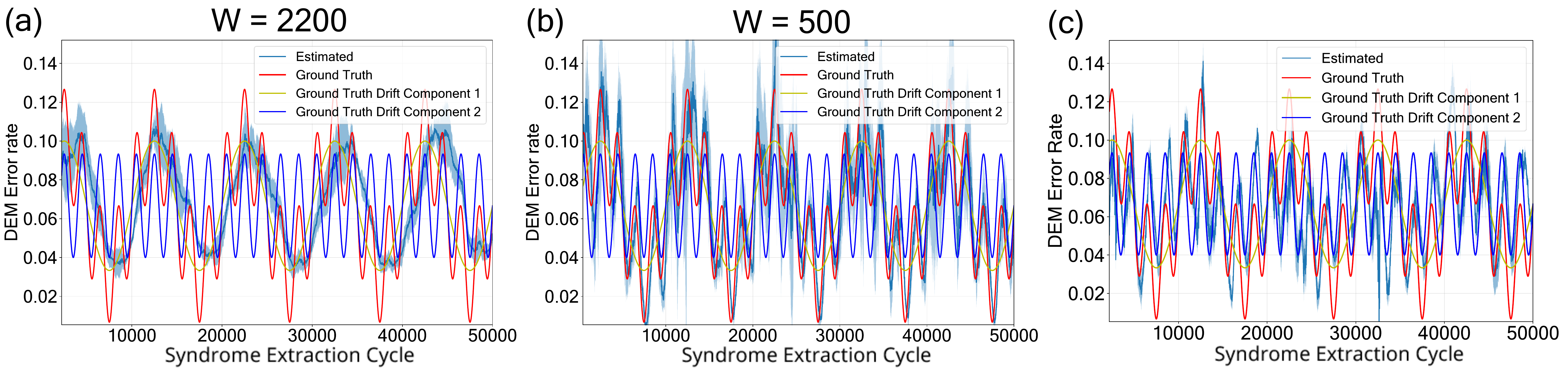}
    \caption{ \color{black}{Low-pass and band-pass filtering behavior of sliding window estimation, for a $d=3$ repetition code under phenomenological noise. All data and ancilla qubits experience uniform depolarizing per syndrome extraction cycle, with time varying error rate $g(t)$ defined by $g_0 = 0.1$, $g_1 = 0.05$, $g_2 = 0.04$, and frequencies  $\omega_1 = 2\pi/(10^4 \Delta t$),  $\omega_2 = 2\pi/(2\times 10^3\Delta t)$.  (a) Estimated DEM error rate of bulk edge (light blue), averaged over all qubits, as a function of syndrome extraction cycle, for a  window size of  $W=2200$. Only the low-frequency component (yellow) $\omega_1 = 2\pi/(10^4 \Delta t$)  is captured, meaning that the sliding window acts as a low-pass filter. The red line shows the full ground-truth signal consisting of two frequencies and the dark blue line shows the more rapidly drifting component. (b) Same as in (a) for a window size of $W=500$, which is able to capture the full time signal. (c) DEM error rate obtained by subtracting the signals of (a) and (b) to isolate the component drifting with frequency $\omega_2 = 2\pi/(2\times 10^3\Delta t)$. }}
    \label{fig:band}
\end{figure*}

Next, we consider the case of a more complicated time-varying signal that consists of two different drift frequencies. We set $g_0 = 0.1$, $g_1 = 0.05$, $g_2 = 0.04$, and frequencies  $\omega_1 = 2\pi/(10^4 \Delta t$),  $\omega_2 = 2\pi/(2\times 10^3\Delta t)$.  The total number of syndrome extraction cycles in this case is set to $5 \times 10^4$. {\color{black}{
We also repeat the simulation five times to plot the error bars. In Fig. \ref{fig:band}(a) we show again the ground-truth (red) and estimated DEM error rate of bulk edge (light blue) as a function of syndrome extraction cycle, for a window size of $W=2200$.
We also show with yellow the first ground-truth frequency component at $\omega_1$, and with dark blue the second ground-truth frequency component at $\omega_2$. The window size we selected captures the frequency component that drifts less rapidly, but we see that the estimated signal experiences a time-lag.  In Fig.~\ref{fig:band}(b), we consider a small window size of $W=500$, which creates a higher uncertainty in the estimated error rate, but captures to an overall good agreement the full time signal of the DEM error rates. Note that with the sliding window method, we cannot resolve the faster component on its own. If we naively subtract the estimated signal of the first component from the full estimated signal, we obtain the signal shown in Fig.~\ref{fig:band}(c). {\color{black}{Note that to construct the estimated signal in Fig.~\ref{fig:band}(c), we have added the estimated $g_0$  component we obtain from the window size $W=2200$, since this approximately the same for the signals of Fig.~\ref{fig:band}(a) and Fig.~\ref{fig:band}(b).}} The signal we obtain in Fig.~\ref{fig:band}(c) is a good approximation of the ground-truth frequency component but still suffers from statistical uncertainty and time-lag features. These arise from the individual estimations using window sizes $W=500$ and $W=2200$, respectively. The statistical uncertainty is large for $W=500$, as per Eq.~(\ref{eq:Stand_deviation}), while the time lag is large for $W=2200$ due to the phase shift induced in the estimation process, described by Eq.~(\ref{Eq:pijOmega}).
\subsection{Application of iterative sliding-window estimation to a phenomenological noise model}
In practical scenarios, we will be completely unaware of the frequencies that exist in the time-signal, and hence we can employ the iterative sliding-window estimation method. We perform a simulation where we consider $10^4$ syndrome extraction cycles. We further set the threshold $\mu =1-\epsilon= 0.22$, and examine two representative test cases. First, we assume a two-frequency noise spectrum with parameters $g_0 = 0.06$, $g_1 = 0.02$, $g_2 = 0.025$ and  frequencies  $\omega_1 = 2\pi/(10^4 \Delta t$),  $\omega_2 = 2\pi/(7\times 10^3\Delta t)$ . These parameters are assumed to be the same for all qubits. The iterative algorithm detailed in Section~\ref{Sec:IterativeSliding} is applied here by varying the window size from $W = 10^4$ down to $W = 10^3$ with a step size of $10^3$. Figure~\ref{fig:iter_window}(a) shows with red line the ground-truth DEM error rate of bulk edge
and with blue line the average estimated DEM error rate of bulk edge . The blue shade corresponds to the standard deviation.
In Fig.~\ref{fig:iter_window}(b) we consider an extended three-frequency spectrum by adding a component $g_3 = 0.015$, with  frequency $\omega_3 = 2\pi/(5 \times10^4 \Delta t$).   The estimated signal in both cases is again obtained by averaging the results of 5 independent estimation trials using the same
parameter. The iterative procedure was terminated at a  window size of $W=1000$, as smaller windows exhibit a higher standard deviation in their estimates, making them less reliable.
In both cases of Fig.~\ref{fig:iter_window} we see good agreement between the estimated and ground-truth error rates. Deviations are attributed to the fact that we use only 10 windows to obtain the estimated DEM error rates, but the accuracy can be improved using a finer step size for selecting window sizes.

\begin{figure*}[!htbp]
    \centering
    \includegraphics[width=1.00\linewidth]{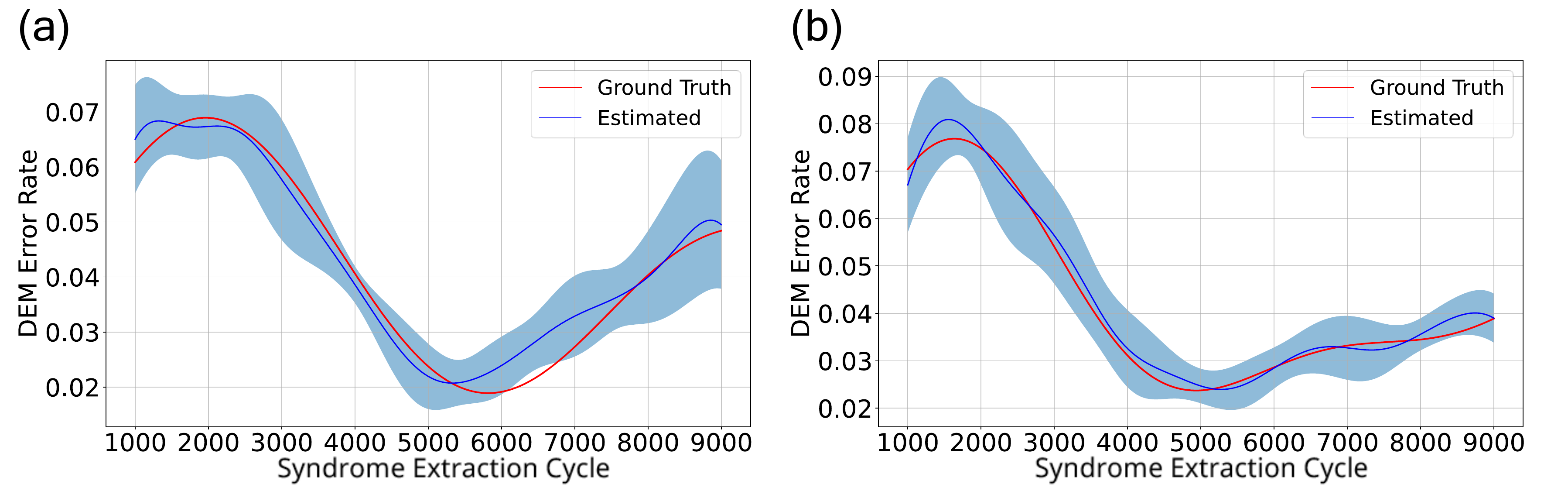}
    \caption{Estimating DEM error rates using iterative sliding window approach with $\mu=0.22$, for a $d=3$ repetition code under phenomenological noise model. Both data and ancilla qubits experience depolarizing noise with time-varying parameters $g(t)$ per syndrome extraction cycle. (a) Estimated DEM error rate of bulk edge (blue), averaged over all qubits, as a function of syndrome extraction cycle in two-frequency noise spectrum. The time-dependent noise, $g(t)$, is defined by parameters $g_0 = 0.06$, $g_1 = 0.02$, $g_2 = 0.025$ with frequencies  $\omega_1 = 2\pi/(10^4 \Delta t$),  $\omega_2 = 2\pi/(7\times 10^3\Delta t)$. (b) Estimated DEM error rate of bulk edge (blue), averaged over all qubits, as a function of syndrome extraction cycle, for a three-frequency noise spectrum. The first two frequency parameters are the same as in (a) with an additional high-frequency component defined by parameters  $g_3 = 0.015$,  $\omega_3 = 2\pi/(5 \times10^4 \Delta t$). In both cases, the ground-truth signal is shown with red curve.}
    \label{fig:iter_window}
\end{figure*}

\subsection{Application of relative window estimation to a phenomenological noise model}
We now discuss our numerical results when using the relative window estimation method, as detailed in Sec.~\ref{Relative Window Estimation}. By rearranging Eq.~\eqref{eq:relative window estimation} we get:
\begin{equation}
p_{ij}(t_l) = W\left(p_{ij,W+1}^{\mathrm{est}}(t_l + \Delta t) - p_{ij,W}^{\mathrm{est}}(t_l)\right) +  p_{ij,W+1}^{\mathrm{est}}(t_l + \Delta t).
\label{eq:diff_eqn}
\end{equation}
The first term in Eq.~\eqref{eq:diff_eqn} represents a discrete differentiation of the estimated probability function in the time domain, scaled by the window size $W$. Since $p_{ij,W}^{\mathrm{est}}(t)$ is a statistical average, it is inherently non-differentiable and exhibits local peaks due to statistical fluctuations. To address this, we apply the Savitzky-Golay \cite{savitzky1964smoothing}  filtering to ensure the functions $p_{ij,W}^{\mathrm{est}}(t)$ and $p_{ij,W+1}^{\mathrm{est}}(t)$ are at least $C^1$ differentiable by fitting polynomial functions to smooth the local peaks.

\begin{figure*}[!ht]
    \centering
    \includegraphics[width=1.02\linewidth]{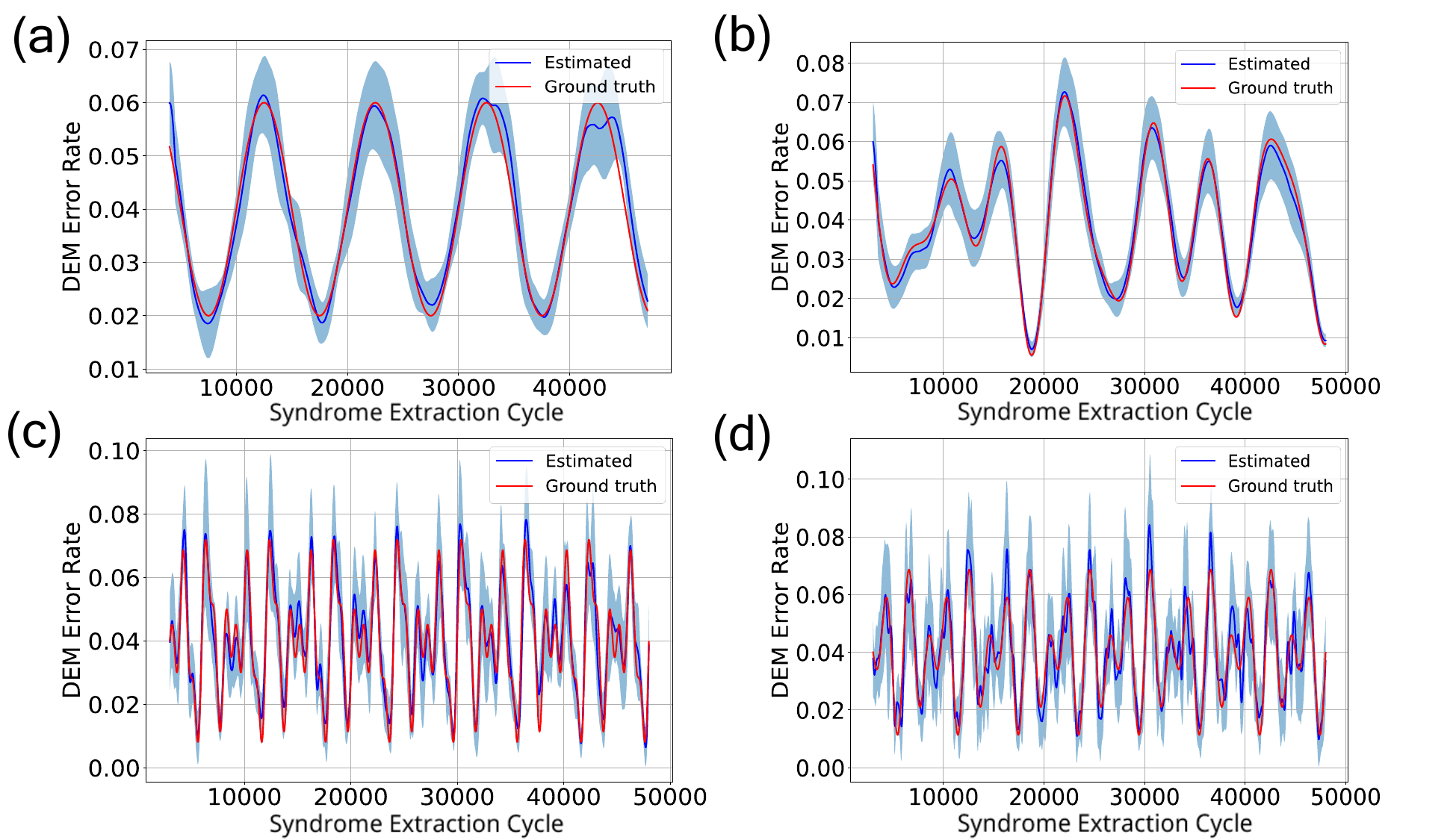}
    \caption{Estimating DEM error rates using the relative window approach for a $d=3$ repetition code under  phenomenological noise model. All data and ancilla qubits experience uniform depolarizing noise with time varying parameter, $g(t)$, per syndrome extraction cycle. (a) Estimated (blue) and ground-truth (red) DEM error rates of bulk edge averaged over all qubits, as a function of syndrome extraction cycle in a single-frequency spectrum. The time-dependent noise, $g(t)$, is defined by parameters $g_0 = 0.06$, $g_1 = 0.03$ with frequency $\omega_1 = 2\pi/(10^4 \Delta t$). (b)  Estimated DEM error rate of bulk edge averaged over all qubits, as a function of syndrome extraction cycle in a three-frequency spectrum. The noise parameters are $g_0 = 0.06$, $g_1 = 0.02$, $g_2 = 0.025$, $g_3 = 0.015$ with frequencies  $\omega_1 = 2\pi/(10^4 \Delta t$), $\omega_2 = 2\pi/(7\times 10^3\Delta t)$ and $\omega_3 = 2\pi/(5\times 10^3\Delta t)$. (c) Estimated DEM error rate of bulk edge averaged over all qubits, as a function of syndrome extraction cycle in a high-frequency dominated spectrum. The amplitudes are the same as in (b), but the frequencies are higher and defined by $\omega_1 = 2\pi/(3 \times 10^3 \Delta t$), $\omega_2 = 2\pi/(2\times 10^3\Delta t)$ and $\omega_3 = 2\pi/(10^3\Delta t)$ (d) Estimated DEM error rate of bulk edge averaged over all qubits, as a function of syndrome extraction cycle in a high-frequency spectrum composed of two frequencies. The noise parameters are $g_0 = 0.06$, $g_1 = 0.02$, $g_2 = 0.025$ and the frequencies are $\omega_1 = 2\pi/(500 \Delta t$) and $\omega_2 = 2\pi/(700\Delta t)$.}
    \label{fig:relative_window_est}
\end{figure*}

We demonstrate the effectiveness of our method through a series of test cases, summarized in Fig.~\ref{fig:relative_window_est}. For all test cases, the overlapping windows used in relative window estimation are $W_1=2000$ and $W_2=2001$ respectively. All panels follow the same convention: the red line represents the ground-truth DEM error rate of bulk edge, and the blue line shows the estimated DEM error rate of bulk edge. The estimated signal is obtained by averaging the results of 5 independent estimation trials.
Figure~\ref{fig:relative_window_est}(a) presents the estimation results for a noise profile with a single drift frequency, parameterized by $g_0 = 0.06$, $g_1 = 0.03$, and  frequency $\omega_1 = 2\pi/(10^4 \Delta t$). Figure~\ref{fig:relative_window_est}(b) shows results for a spectrum with three frequencies, using parameters $g_0 = 0.06$, $g_1 = 0.02$, $g_2 = 0.025$, $g_3 = 0.015$ and  frequencies  $\omega_1 = 2\pi/(10^4 \Delta t$), $\omega_2 = 2\pi/(7\times 10^3\Delta t)$ , $\omega_3 = 2\pi/(5\times 10^3\Delta t)$. In both cases, the estimated error rates approximate well the ground-truth signal. We further test the method's robustness in the case of rapidly drifting noise profile in Fig.~\ref{fig:relative_window_est}(c), where we define the parameters $g_0 = 0.06$, $g_1 = 0.02$, $g_2 = 0.025$, $g_3 = 0.015$ and  frequencies  $\omega_1 = 2\pi/(3 \times 10^3 \Delta t$), $\omega_2 = 2\pi/(2\times 10^3\Delta t)$ , $\omega_3 = 2\pi/(10^3\Delta t)$.  Despite the high-amplitude, rapid oscillations, our estimation method continues to perform well. Finally, Fig.~\ref{fig:relative_window_est}(d) presents an even more challenging case where much higher frequencies ($\omega_1 = 2\pi/(500 \Delta t$), $\omega_2 = 2\pi/(700\Delta t)$) cause the ground truth error rate to drift significantly  on a timescale of only a few syndrome extraction cycles, parameterized by $g_0 = 0.06$, $g_1 = 0.02$, $g_2 = 0.025$. Here, the estimated error rate still matches the ground truth at most time instances; however the standard deviation of the estimate increases noticeably at the local maxima and minima. In summary, the relative window estimation method proves capable of accurately reconstructing a wide variety of noise drift profiles, from slow, gradual drifts to rapid, high-amplitude oscillations.

\subsection{Estimating circuit-level noise}
\label{Estimating circuit-level noise}
\begin{figure*}[!htbp]
    \centering
    \includegraphics[width=1.00\linewidth]{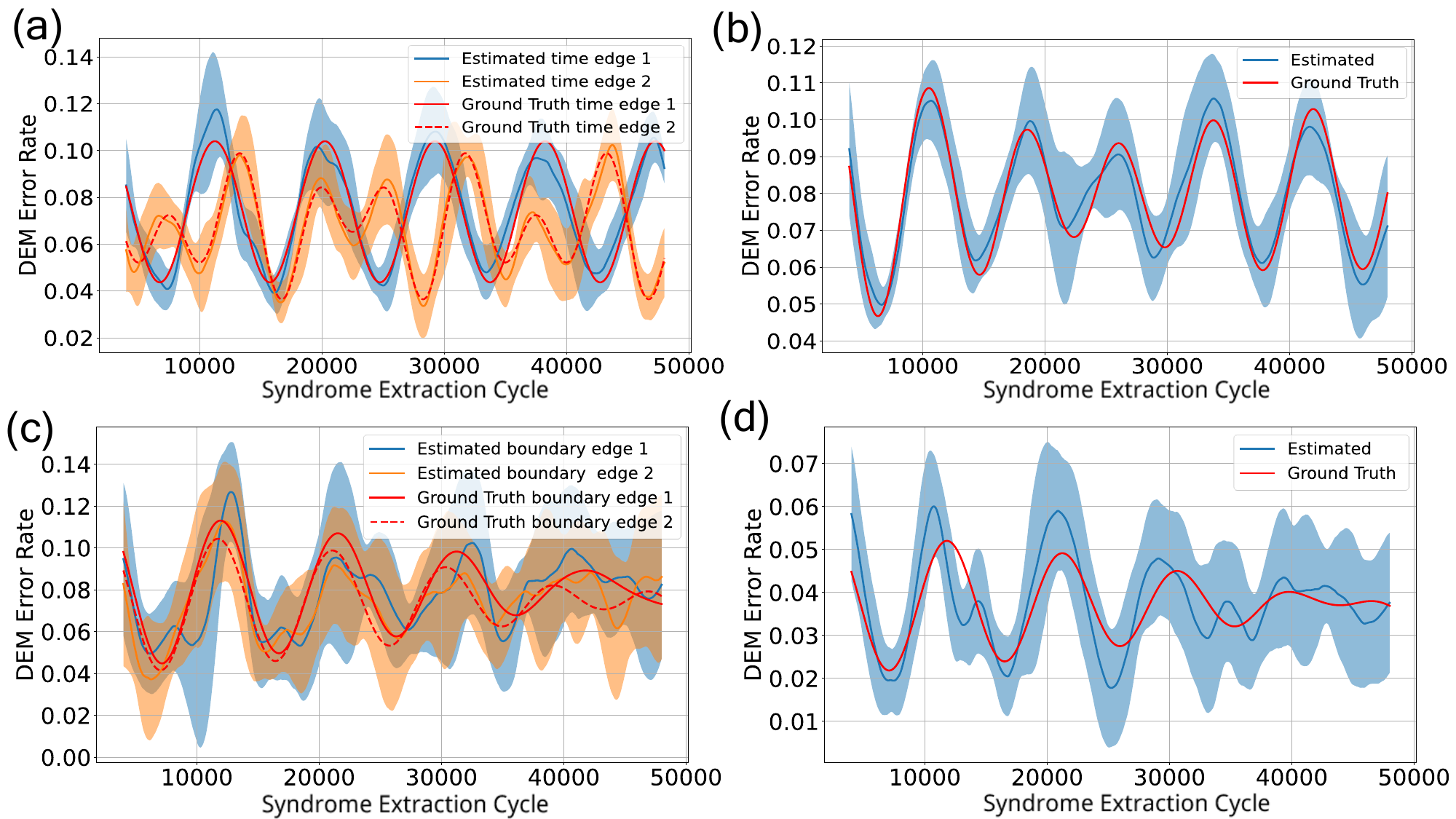}
    \caption{Estimating DEM error rates with the relative window approach, for a  $d=3$ repetition code under circuit-level noise  with
    inhomogeneous error rates for qubits and gates defined in Table~\ref{tab:gt_parameters}. (a) Estimated and ground truth DEM error rates for the time-like edges. (b) Estimated and ground truth DEM error rates for bulk space edge. (c) Estimated and ground truth DEM error rates for the boundary space edges. (d) Estimated and ground truth error rates for diagonal edge. }
    \label{fig:cl_noise_est}
\end{figure*}

To test our methods against a more complete noise model where gates can also fail, we assume a circuit-level noise model. For simplicity we assume that all ancilla and data qubits, as well as the gates experience single-frequency noise drift. However, we change our previous setup and assume non-uniform parameters $g(t)$ for qubits and gates. Ancilla and qubit errors appear as horizontal and vertical edges in the DEM, whereas non-zero gate errors give rise to diagonal edges in the DEM. For $d=3$ repetition code memory that we consider again, there are four CNOTs per round which are used to extract the syndromes of $Z_1Z_2$ and $Z_2Z_3$ stabilizers. The two-qubit errors are modeled as two-qubit depolarizing noise. We assume that the $g_0$ and $g_1$ parameters are the same for all CNOTs. We set a drift frequency for the first two CNOT errors to be the same that measure $Z_1Z_2$, and another drift frequency for the last two CNOT errors that measure $Z_2Z_3$. All the noise parameters that we set for qubits and gates are summarized in Table~\ref{tab:gt_parameters}. 

We perform a simulation of $5 \times 10^4$ syndrome extraction cycles and use the relative window estimation method with two overlapping windows of size $W_1=2000$ and $W_2=2001$ to estimate all types of edges. Our results are shown in Fig.~\ref{fig:cl_noise_est}. We repeat the estimation experiment five times using the same parameters, in order to get the average behavior and show the standard deviation. In each panel of Fig.~\ref{fig:cl_noise_est}, we show the standard deviation as a shaded region, the average of our estimated error rate with a blue or orange line, and the ground-truth signal with red line. 
\begin{table}[htbp!]
\centering
\caption{$g(t)$ parameters used for the circuit-level noise model for a $d=3$ repetition code in Fig.~\ref{fig:cl_noise_est}. Amplitudes ($g_0$, $g_1$) and drift frequencies ($\omega_1$) are shown for each data qubit (q.), and each ancilla qubit (q.), as well as the error parameters for the CNOT gates.}
\label{tab:gt_parameters}
\renewcommand{\arraystretch}{1.3} 
\begin{tabular}{|c|c|c|c|}
\hline
\textbf{Qubit/Gate} & \textbf{$g_{0}$} & \textbf{$g_{1}$} & \textbf{$\omega_{1}$} \\
\hline
Data Qubit $d_{1}$ & 0.07  & 0.035 & $2\pi/(10^4 \Delta t)$ \\
\hline
Data Qubit $d_{2}$ & 0.07  & 0.035 & $2\pi/(8 \times 10^{3}\Delta t)$ \\
\hline
Data Qubit $d_{3}$ & 0.06  & 0.03  & $2\pi/(9 \times 10^{3}\Delta t)$ \\
\hline
Ancilla Qubit $a_{1}$ & 0.04  & 0.025 & $2\pi/(9 \times 10^{3}\Delta t)$ \\
\hline
Ancilla Qubit $a_{2}$ & 0.04  & 0.03  & $2\pi/(6 \times 10^{3}\Delta t)$ \\
\hline
CNOT errors for $Z_1Z_2$ & 0.045 & 0.03 & $2\pi/(9 \times 10^{3}\Delta t)$ \\
\hline
CNOT errors for $Z_2Z_3$ & 0.045 & 0.03 & $2\pi/(10^{4}\Delta t)$ \\
\hline
\end{tabular}
\end{table}
Figure~\ref{fig:cl_noise_est}(a) presents the estimated error rate for horizontal (timelike) edges in the decoding graph. For a $d=3$ repetition code, which utilizes two ancilla qubits, there are two corresponding timelike edges per syndrome extraction cycle. The estimated error rates for these edges, shown in blue and orange, demonstrate strong agreement with the ground truth values. Figure~\ref{fig:cl_noise_est}(b) shows the estimated error rate for the unique vertical bulk (spacelike) edge in the same code. The estimated value, in blue, tracks the ground-truth, in red, consistently, with deviations observed mainly at local maxima. Figure~\ref{fig:cl_noise_est}(c) displays the results for the two vertical boundary edges. The boundary edge error rates are calculated using Eq.~\eqref{eq:bound_edge}. The uncertainty from estimated bulk edge error rates propagates to the boundary estimates, resulting in a more pronounced statistical uncertainty for boundary edges. Finally, Fig.~\ref{fig:cl_noise_est}(d) presents the unique diagonal edge, which arises from gate errors in the spacetime graph. The ground-truth error rate for this edge, shown in red, exhibits a complex, multi-frequency oscillation with a decaying envelope. This behavior is a direct result of the interference between the independently drifting noise parameters of the ancilla qubits, data qubits, and gates. While the estimated error rate, shown in blue, successfully captures the overall decaying oscillatory trend, mismatches occur due to the estimator's inability to fully capture the high-frequency components which cause decay in the ground-truth signal.

\subsection{Decoding Results \label{Decoding Results}} 
We now turn our attention to the the decoding performance under drifting noise. Here, we will compare the decoding performance of our estimated DEM against the ground-truth DEM. Since our estimation methods track drifting noise across a large number of syndrome extraction cycles in the same QEC experiment, we adopt the following approach to calculate the logical error rate. After learning the error rates and reconstructing the estimated DEM, we decode both ground-truth and estimated DEMs for some specified number of syndrome extraction cycles, using the same detection events to compare them fairly. To quantify the logical error mismatch, we define the relative logical error rate, $\Delta$, as:
\begin{equation}
    \Delta = \left(\frac{\epsilon_L^{\text{est.}}}{\epsilon_L^{\text{stim}}}\right) - 1
\end{equation}
where $\epsilon_L^{\text{est.}}$ is our estimated logical error rate per cycle, and $\epsilon_L^{\text{stim}}$ represents the logical error rate per cycle obtained by decoding the ground-truth DEM.

\begin{figure*}[!htbp]
    \centering
    \includegraphics[width=1.0\linewidth]{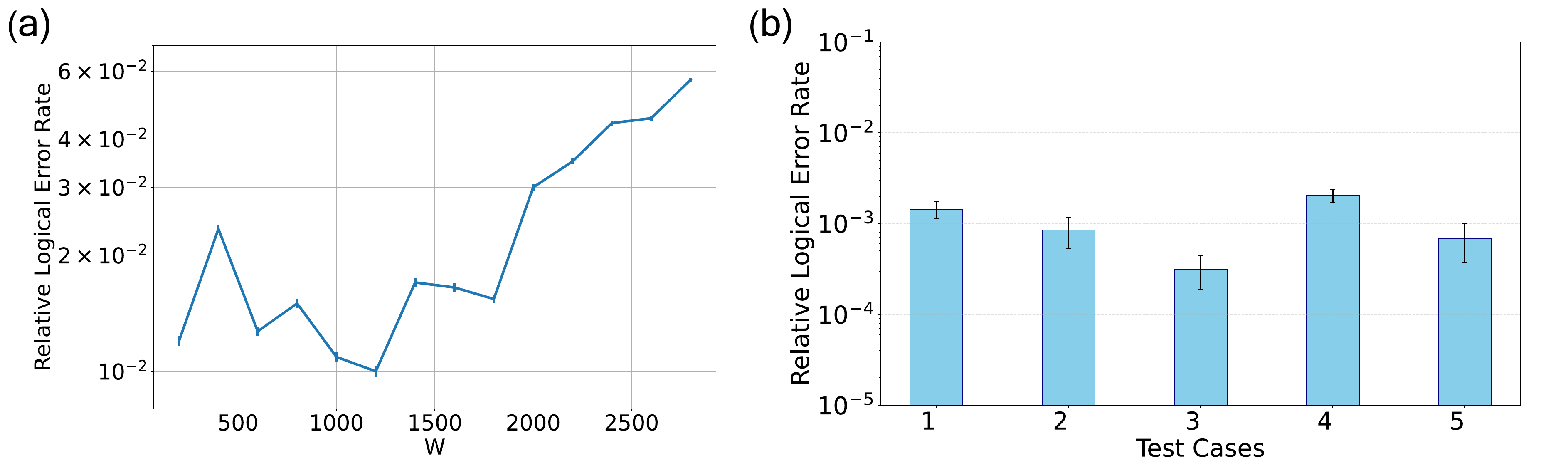}
    \caption{ Relative logical error rate $|\Delta|$, defined as the relative error between the logical error rate obtained from the estimated DEM and the one obtained from the ground-truth DEM.  (a) $|\Delta|$ versus window size for a $d=3$ repetition code under a phenomenological noise model. The time-varying noise parameters are uniform across all data and ancilla qubits, defined by $g_0 = 0.1$, $g_1 = 0.05$, and a single drift frequency $\omega_1 = 2\pi/(10^4 \Delta t$). To estimate the DEM we use the sliding window method and a fixed window size for each data point. (b) $|\Delta|$ across five distinct test cases. Test cases 1--4 correspond to estimations performed under the phenomenological noise model, using the specific parameter sets detailed in Fig.~\ref{fig:relative_window_est}(a)--(d), respectively. Test case 5 corresponds to estimation under a circuit-level noise model, with parameters as described in Fig.~\ref{fig:cl_noise_est}.}
    \label{fig:rler}
\end{figure*}

We consider the single-pass sliding-window estimation and analyze how the error metric $\Delta$ varies with the window size. We consider a phenomenological noise model for a $d=3$ repetition code memory with uniform parameters for the ancilla and data qubits, selected as $g_0=0.1$, $g_1=0.05$, and a single drift frequency $\omega_1 = 2\pi/(10^4 \Delta t$). We further use $n=10^3$ cycles and $5 \times 10^4$ shots for the logical error rate calculation. In Fig.~\ref{fig:rler}(a) we show the logical error rate as a function of the window size.  Larger window sizes introduce greater estimation errors due to the pronounced damping factors and time-lags predicted by our spectral analysis leading to a larger mismatch between the estimated and ground-truth logical error rates. In the opposite regime, very small windows lead to increased statistical uncertainty in the error rates. Also, we note that  $|\Delta|$ attains higher values for larger window size in comparison to smaller window sizes. This shows that  time-lag and damping errors might be more detrimental compared to statistical uncertainty errors. The trade-off between these competing features is balanced by a clear minimum at $W=1250$, matching very closely to our theoretically predicted optimum of $W_{\text{opt}} = 1228 \pm 42$ when calculated from Eq.~\eqref{eq:transf_fun} with $\epsilon = 0.05$. The close agreement between theory and numerical simulation demonstrates the effectiveness of our window-based optimization framework and how it can 
inform decoders about noise drifts, leading to optimal decoding performance.

As a next example, we evaluate the decoding performance under complex noise spectra by analyzing test cases containing multiple frequency components. We now consider the case where the error rates are obtained based on the relative-window estimation method. Figure~\ref{fig:rler}(b) presents the results for both phenomenological and circuit-level noise models, with test cases we previously presented in Figs.~\ref{fig:relative_window_est},~\ref{fig:cl_noise_est}. Using again $10^3$ syndrome extraction cycle and $5 \times 10^4$ shots for the logical error rate calculation, we observe that our relative window estimation method can achieve remarkable relative precision in the logical error rate to the order of $\sim 10^{-4}-10^{-3}$. This performance persists even under cases when significant drift in noise occurs within just a few syndrome extraction cycles, demonstrating the robustness of our approach. The small magnitude of $\Delta$ indicates that our estimation technique successfully tracks the time-varying error rates while maintaining decoding accuracy comparable to ground-truth knowledge of the noise parameters.

\begin{figure*}[!htbp]
    \centering
    \includegraphics[width=1.0\linewidth]{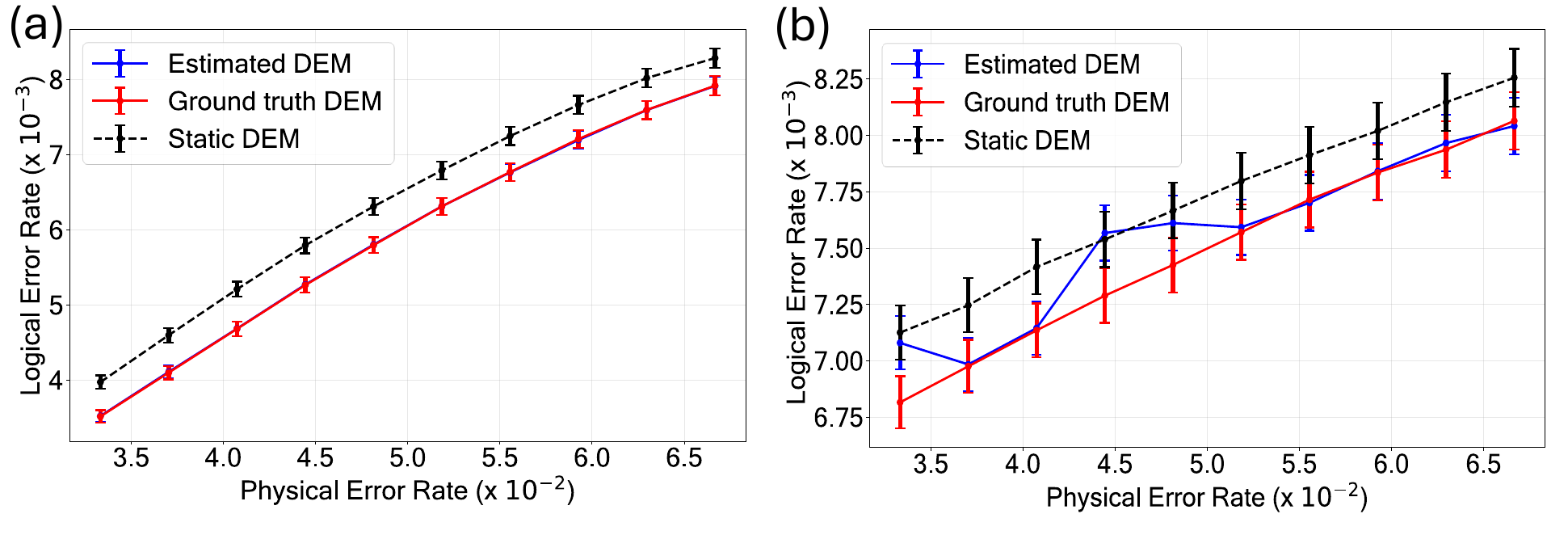}
    \caption{ Comparing the logical performance of estimated versus ground-truth and static DEMs, for a $d=3$ repetition code.
     (a) Logical error rate as a function of the physical error rate for a  $d=3$ repetition in phenomenological noise model, with parameters $g_0 = g_*$ and $g_1 = g_*/2$, where $g_*$ (physical error rate) is varied. The blue line corresponds to the estimated DEM, the red line to the ground-truth DEM, and the black line to the static DEM. The curve for the estimated DEM overlaps with the one of the ground-truth DEM.
     (b) 
     Logical error rate of estimated, ground-truth and static DEMs as a function of the physical error rate in circuit-level noise model. The ancilla and gate error parameters are defined in Table~\ref{tab:gt_parameters}. The physical error rate that is varied is the parameter $g_*=g_0=2g_1$ of the data qubits, whereas the drift frequency of data qubits is found in Table~\ref{tab:gt_parameters}.}
    \label{fig:ler_vs_per}
\end{figure*}

 We now analyze the behavior of the logical error rate per cycle ($\ell$) across different physical error rates for both phenomenological and circuit-level noise models in the repetition code. The estimation in both noise models is done using relative window estimation with two overlapping windows of size $2000$ and $2001$ respectively.  In the phenomenological noise model, the time-varying error probability $g(t)$ has a single-frequency component with parameters $g_0 = g_*$, $g_1 = g_*/2$, and frequency $\omega_1 = 2\pi/(10^4 \Delta t$). {\color{black}{For simplicity, we assume the same error rates for all qubits.}}  The physical error rate, $g_*$, is varied to study its effect on the logical error rates. For each configuration, we use $10^2$ cycles and $5 \times 10^5$ shots to compute the logical error rates. Figure~\ref{fig:ler_vs_per}(a) compares the logical error rate performance of three different decoders. The first decoder is adapted using our estimated DEM (blue line), the second decoder uses ground-truth DEM (red line), and the third uses a static DEM that assumes no drift (black line). The static DEM is configured with the parameters $g_0 = g_*$ and $g_1 = 0$. The results demonstrate that our adaptive decoder consistently achieves logical error rates that match the ground truth across the entire range of physical error rates. In contrast, the decoder based on the static DEM exhibits significantly higher logical error rates. This comparison further validates the robustness of our method in effectively tracking time-varying noise, regardless of the underlying physical error rate.

Figure~\ref{fig:ler_vs_per}(b) presents the logical error rate performance for of $d=3$ repetition code in circuit level noise model with decoders utilizing our estimated DEM, the ground-truth DEM, and a static DEM (configured with $g_0 = g_*$, $g_1 = 0$).The decoder using our estimated DEM maintains good agreement with the ground truth, though slight deviations are observed at certain physical error rates. These deviations are primarily due to increased statistical uncertainty when estimating the boundary and diagonal edge weights, as discussed in Section~\ref{Estimating circuit-level noise}. The decoder with the static DEM again exhibits a higher logical error rate compared to the decoder based on our estimated DEM. 

\begin{table}[htbp!]
\centering
\caption{Drift frequencies used for the phenomenological noise model in Fig.~\ref{fig:ler_vs_per_sc}, for a $d=3$ $X$-memory rotated surface code. Drift frequencies are shown for each data and ancilla qubit. }
\label{tab:sc_N1_parameters}
\renewcommand{\arraystretch}{1.3} 
\begin{tabular}{|c|c|c|c|}
\hline
\textbf{Qubit} & \textbf{$\omega_{1}$} \\
\hline
Data Qubit $d_{1}$  & $2\pi/(5800 \Delta t)$ \\
\hline
Data Qubit $d_{2}$  & $2\pi/(9800\Delta t)$ \\
\hline
Data Qubit $d_{3}$   & $2\pi/(4800\Delta t)$ \\
\hline
Data Qubit $d_{4}$   & $2\pi/(8800\Delta t)$ \\
\hline
Data Qubit $d_{5}$   & $2\pi/(12800\Delta t)$ \\
\hline
Data Qubit $d_{6}$   & $2\pi/(7800\Delta t)$ \\
\hline
Data Qubit $d_{7}$   & $2\pi/(11800\Delta t)$ \\
\hline
Data Qubit $d_{8}$   & $2\pi/(6800\Delta t)$ \\
\hline
Data Qubit $d_{9}$   & $2\pi/(10800\Delta t)$ \\
\hline
Ancilla Qubit $a_{1}$  & $2\pi/(5800\Delta t)$ \\
\hline
Ancilla Qubit $a_{2}$  & $2\pi/(9800\Delta t)$ \\
\hline
Ancilla Qubit $a_{3}$  & $2\pi/(4800\Delta t)$ \\
\hline
Ancilla Qubit $a_{4}$  & $2\pi/(8800\Delta t)$ \\
\hline
\end{tabular}
\end{table}

{\color{black}{Now we present the logical error rate per cycle  across different physical error rates for a $d=3$ rotated surface code in phenomenological noise model. In this case, we measure only the X-stabilizers and the corresponding X-DEM is decoded to get the logical error rate. The time-varying parameters are defined as $g_0 = g_*$, $g_1 = g_*$, which are applied uniformly to all data and ancilla qubits. The drift frequencies for data and ancilla qubits are specified in Table~\ref{tab:sc_N1_parameters}. The physical error rate, $g_*$, is varied to study its effect on the logical error rates. Estimation is again done using the relative window estimation method with two overlapping windows of size $2000$ and $2001$, respectively. We use $50$ cycles and $10^6$ shots to compute the logical error rates. Figure~\ref{fig:ler_vs_per_sc} shows the logical error rate when the decoder uses our estimated DEM (blue), the ground-truth DEM (red), or the static DEM (black). The static DEM is again configured with the parameters $g_0 = g_*$ and $g_1 = 0$. The results show that the decoder adapted using our estimated DEM achieves improved logical error suppression relative to the static DEM, which assumes stationary error rates. Furthermore, the estimated DEM yields logical error rates that closely match those obtained with the ground-truth DEM across the full range of physical error rates.}}

It is important to note that these logical error rates were calculated using only $\sim 10^2$ cycles. Given that the ground-truth error rate drifts with a frequency of $2\pi/(10^4 \Delta t)$, we anticipate that the performance gap between the static and adaptive decoders will become even more pronounced when cycles on the order of $10^3$ or $10^4$ is used for benchmarking. Overall, our adaptive decoder improves decoding performance  over static decoders by consistently matching the ground-truth logical error rates. 
\begin{figure*}[!htbp]
    \centering
    \includegraphics[width=0.5\linewidth]{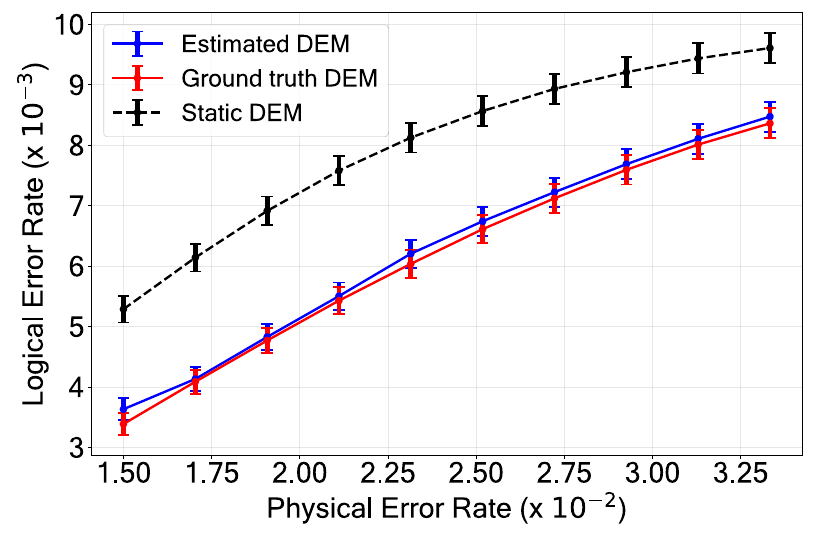}
    \caption{ Comparing the logical performance of estimated versus ground-truth and static DEMs, for a $d=3$ rotated surface code in phenomenological noise model. The logical error rate is calculated using only the X-type stabilizers, and decoding is performed using the X-DEM. The blue line shows the performance of the estimated DEM, the red line the performance of the ground-truth DEM, and the black line shows the performance of the static DEM. The error rates are set to $g_0 = g_*$ and $g_1 = g_*$, where $g_*$ (the physical error rate) is varied uniformly for all data and ancilla qubits. The drift frequencies for all qubits are provided in Table~\ref{tab:sc_N1_parameters}.}
    \label{fig:ler_vs_per_sc}
\end{figure*}

\section{Discussion} \label{Discussion}
Our results show that the proposed estimation and decoding methods remain robust across a wide range of error rates.  Within realistic regimes, the sliding-window framework provides both accuracy and adaptability, making it a promising tool for scalable characterization of QEC experiments. It is further possible to extend the sliding-window analysis to overlapping windows that move backwards in time  intervals $[t -( W + 1)\Delta t, t)$ to estimate errors for early cycles, thereby enabling full estimation of arbitrarily long experiments. Additionally, our framework generalizes naturally to consecutive experimental runs, where noise may vary across shots, instead of syndrome extraction cycles we considered in this work.

Other works have developed methods for tracking noise drift; however, these approaches come with certain limitations. Reference~\cite{Wang2023DGR} addresses the issue of noise drift 
by tracking the occurrence of edges from matchings and the correlations between edges obtained from multiple experiments. This is done via heuristic methods which, as the authors mention, may prove ineffective in cases where edges associate with numerous correlated counterparts. Additionally, Ref.~\cite{Wang2023DGR} employs neural network models that require gradient-based optimization. Such methods often depend on the initial guess and might require parameter tuning to obtain good solutions. In another work, the authors of Ref.~\cite{kobori2024bayesian} used low-rank tensor approximations and Monte Carlo methods to estimate time-dependent noise from syndromes. However, the low-rank approximation causes systematic bias in the estimation and requires optimization of hyperparameters to achieve accuracy in the estimation. Further, the tensor network simulation assumes that the syndrome extraction is noise-free, hindering the applicability of this method to realistic noise scenarios. {\color{black}{Reference~\cite{PhysRevApplied.9.064042} presents a machine-learning framework employing kalman and regression-based filters to predict qubit dynamics under dephasing noise, each exhibiting trade-offs in accuracy and robustness. Among all these filters, the autoregressive kalman filter performs best in prediction, but it requires extensive pre-training. Other filters exhibit reduced robustness to measurement noise and poor forward prediction. Overall, while these methods offer a solution to tracking time-dependent noise, the trade-offs impose constraints on their applicability in general noise settings.}} Finally, Ref.~\cite{huo2017learning} utilizes Gaussian process regression methods to predict time-dependent error rates, but the methods they propose are approximate, leaving an open question as to whether these methods can be accurate enough in general noise settings.

Looking forward, there are several avenues that can be explored based on our proposed methods. Building on the computational efficiency of our analytical framework, incorporating advanced pre- and post-processing techniques such as Kalman filtering \cite{kalman1960new} could further improve the ability to track rapid drift and high-frequency fluctuations. Experimental validation will be an essential next step, particularly for testing adaptive estimation and decoding across different platforms and for various QEC codes. Our framework can also be applied to the case of hypergraph DEMs, as the only extension necessary to our framework is to track multi-point correlations instead of up to two-point correlations that we used in this work.  Finally, extending the methodology to non-Markovian noise regimes offers an intriguing direction, as it would test the limits of window-based estimation under more complex temporal correlations.

\section{Conclusion\label{Conclusion}}
Syndrome data from QEC experiments provide a rich source of information which can benefit decoders. Noise characterization based on the syndrome history is particularly useful in scenarios where conventional methods such as tomography become impractical, especially for large devices. The ability to track noise drift during QEC experiments is critical for understanding and improving the performance of error-corrected architectures. Our framework demonstrates that we can reliably estimate drifting error rates to inform decoders. Our sliding window estimation methods enable frequency filtering, providing a valuable insight into individual frequency components that are present in real devices. Additionally, our relative window estimation method allows us to estimate error rates in the more challenging scenario of rapid drifts. Remarkably, our methods incur no additional experimental cost, besides the syndrome collection, and very low computational sources.  The techniques we introduced  establish a foundation for adaptive decoding strategies that respond in real time to evolving environments. Our framework points toward a new paradigm where noise is not only corrected but also continuously monitored and understood during QEC experiments.

\acknowledgements{The authors acknowledge support from the Office of
the Director of National Intelligence (ODNI), Intelligence
Advanced Research Projects Activity (IARPA), under
the Entangled Logical Qubits program through Cooperative Agreement Number W911NF-23-2-0216 and the
ARO/LPS QCISS program (W911NF-21-1-0005).}

\appendix 
\section{Derivation of Estimated Error rates using Sliding Window} \label{Appendix A}
\begin{figure*}[!htbp]
    \centering
    \includegraphics[width=1.01\linewidth]{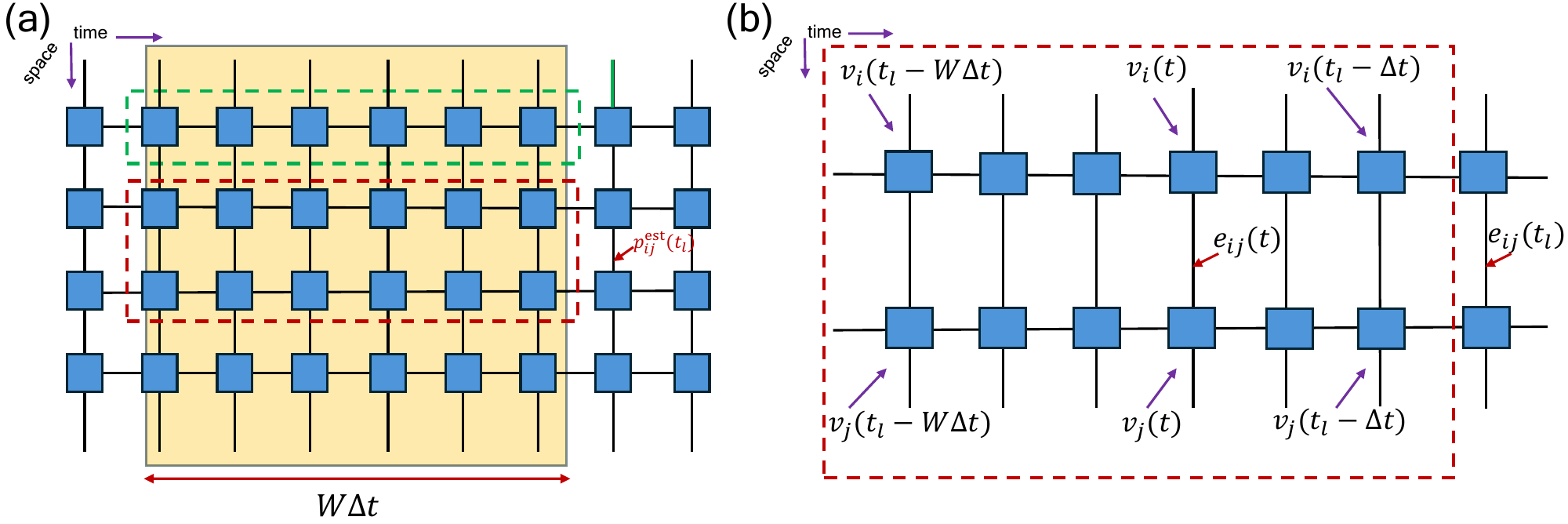}
    \caption{Schematic of the sliding window method on the space-time graph. (a) A window of fixed duration $W\Delta t$ slides along the temporal axis. The error rate $p_{ij}^{\text{est}}(t_l)$ at time $t_l$ is estimated from the syndrome data within the interval $t \in [t_l - W\Delta t, t_l)$. (b) Detailed view of the window contents, showing the dynamical vertex state $v_i(t)$ and an edge $e_{ij}(t)$ over $W$ time steps, which are used to estimate error rate $p_{ij}^{\text{est}}(t_l)$. }
    \label{fig:sliding_window}
\end{figure*}

Here we explain in detail how Eq.~(\ref{eq:bulk_edge}) and Eq.~(\ref{eq:bound_edge}) in the main text are adapted to capture time-varying noise. We consider a time window of width, $W$, defined by $t\in [t_l-W\Delta t, t_l)$, and focus on the edge state $e_{ij}(t)$ {\color{black}{[we follow the same notation as in Ref.~\cite{Spitz2018}]}}.  This edge state is a binary variable; when it attains the value 1, it denotes the error event that flips vertices $v_i(t)$ and $v_j(t)$ at time $t$. We record the outcome of $e_{ij}(t)$ across syndrome extraction cycles in the window, $W$, into the bitstring:
\begin{equation*}
    B_W = b_0b_1b_2\dots b_{W-1},
\end{equation*}
where
\begin{equation*}
    b_k=e_{ij}(t_0 - W\Delta t + k\Delta t), ~~k\in[0,W-1]
\end{equation*}
is the instantaneous value of $e_{ij}(t)$  at some particular time within the time window (see Fig.~\ref{fig:sliding_window}).

The expectation value, $\langle e_{ij}(t) \rangle$, over time interval $t\in [t_l-W\Delta t, t_l)$, can then be obtained from:
\begin{equation}
\langle e_{ij}(t) \rangle = \frac{1}{W}\sum_{k=0}^{W-1} b_k,
\label{eq:window_estimate_edge}
\end{equation}
where the sum counts the number of 1s in the string, with $n \in \{0,1,\dots,W\}$.

{\color{black}{
Under the assumption of statistical independence between error events, the expectation value of a bulk edge:
\begin{equation}
\langle e_{ij} \rangle = \frac{1}{2} - \sqrt{\frac{1}{4} - \frac{\langle v_i v_j \rangle - \langle v_i \rangle\langle v_j \rangle}{1 - 2(\langle v_i \rangle + \langle v_j \rangle) + 4\langle v_i v_j \rangle}},
\label{eq:bulk_edge_append}
\end{equation}
and the expectation value of a boundary edge:
\begin{equation}
\langle e_{ii} \rangle = \frac{1}{2} + \frac{\langle v_i \rangle - 1/2}{\prod_{j \ne i} (1 - 2p_{ij})},
\label{eq:bound_edge_append}
\end{equation}
can be used for both static or drifting error rates \cite{Spitz2018}  
}}. In the case of static noise model, $\langle e_{ij}\rangle = p_{ij}$ and $\langle e_{ii}\rangle =p_{ii}$. We then arrive at Eq.~(\ref{eq:bulk_edge}) and Eq.~(\ref{eq:bound_edge}).

When the estimation under noise drift is done using Eq.~(\ref{eq:bulk_edge_append}),  (\ref{eq:bound_edge_append}) a crucial question arises: what information does $\langle e_{ij}(t) \rangle$, computed over a finite window $W$, capture about the true error rate $p_{ij}(t)$? In the following, we denote  $p_{ij}^{\mathrm{est}}(t_l)=\langle e_{ij}(t) \rangle$ and subsequently show how $p_{ij}^{\mathrm{est}}(t_l)$ is related to the true error rate $p_{ij}(t)$.
Using Eq.~\eqref{eq:window_estimate_edge}, $\langle e_{ij}(t) \rangle$ is related to the  number of 1s in the bitstring $B_W$, defined by $n = \sum_{k=0}^{W-1} b_k$ . Under drifting noise, $n$ is a random variable. The randomness in $n$ originates from the probabilistic occurrence of the bitstring $B_W$. This probabilistic ocurrence of bitstring captures the information about true error rate $p_{ij}(t)$. Formally the probability of getting a bitstring $B_W$ relates to the true error rates $p_{ij}(t)$ as:
\begin{equation}
\begin{split}
P(B_W) &= \\&\prod_{k=0}^{W-1} \Big[ p_{ij}(t_l - W\Delta t + k\Delta t)^{b_k}\cdot
\\& \left(1 - p_{ij}(t_l - W\Delta t + k\Delta t)\right)^{1 - b_k} \Big],
\label{eq:bit_string_prob}
\end{split}
\end{equation}
where $b_k \in \{0,1\}$ .

The occurrence of a specific bit string $B_W$ among all possible $2^W$ configurations is a stochastic event with probability $P(B_W)$.
{\color{black}{Using the central limit theorem, the expectation value of $n$ (number of `1`s), scaled by a factor of $1/W$ in a sufficiently large window of size $W$, equals $\langle e_{ij}(t) \rangle$  and $p_{ij}^{\mathrm{est}}(t_l)$ .}} The expectation value of $n$ in a window of size $W$ is given by:
\begin{equation}
\begin{split}
    \mathbb{E}[n] &= \sum_{i=0}^{2^W-1} 
    n_{B_W^{(i)}} \cdot P(B_W^{(i)}),
\label{eq:expectation value of n}
\end{split}
\end{equation}
where $B_W^{(i)}$ represents the $i$-th possible bit string of length $W$, $n_{B_W^{(i)}} \in \{0,1,\dots,W\}$ counts the `1`s in $B_W^{(i)}$, and $P(B_W^{(i)})$ is the probability of observing $B_W^{(i)}$. Substituting the expression for $P(B_W^{(i)})$ we get
\begin{equation}
\begin{split}
&\mathbb{E}[n] = \sum_{i=0}^{2^W-1} n_{B_W^{(i)}} \cdot \quad \Big( \prod_{k=0}^{W-1} p_{ij}(t_l- W\Delta t + k\Delta t)^{b_k^{(i)}}\cdot
\\&
(1-p_{ij}(t_l- W\Delta t + k\Delta t))^{1-b_k^{(i)}} \Big).
\end{split}
\end{equation}

We proceed by mathematical induction to generalize the structure for $\mathbb{E}[n]$ to arbitrary window sizes. Assuming the expectation $\mathbb{E}[n_W]$ is known for window size $W$ under the induction hypothesis, we derive $\mathbb{E}[n_{W+1}]$ for the extended window by incorporating the additional time step $t_l$. 

For a window of size $W+1$, the $2^{W+1}$ possible bit strings emerge from augmenting each existing $W$-length string $B_W$ with an additional bit $b_{W} \in \{0,1\}$. Formally, the extended string space $\mathcal{B}_{W+1}$ is generated through the Cartesian product $\mathcal{B}_W \times \{0,1\}$, where each $B_W \in \mathcal{B}_W$ branches into two new configurations: $(B_W,0)$ and $(B_W,1)$.

The addition of a new bit increments each existing count $n_{B_W^{(i)}}$ by either 0 or 1. Consequently, the expectation for window size $W+1$ decomposes as:
\begin{equation}
\begin{split}
\mathbb{E}[n_{W+1}] = & \\
& \sum_{i=0}^{2^W-1} \Big[ (n_{B_W^{(i)}}+1) \cdot P(B_W^{(i)}) \cdot p_{ij}(t_l) \\
& \quad + n_{B_W^{(i)}} \cdot P(B_W^{(i)}) \cdot (1-p_{ij}(t_l)) \Big]
\end{split}
\end{equation}

where the first term accounts for bit strings ending with 1, and the second for those ending with 0.

Simplifying, we obtain:
\begin{equation}
\mathbb{E}[n_{W+1}] = \underbrace{\sum_{i=0}^{2^W-1} n_{B_W^{(i)}} P(B_W^{(i)})}_{\mathbb{E}[n_W]} + p_{ij}(t_l) \underbrace{\sum_{i=0}^{2^W-1} P(B_W^{(i)})}_{1},
\end{equation}
yielding the recursive update as
\begin{equation}
\mathbb{E}[n_{W+1}] = \mathbb{E}[n_W] + p_{ij}(t_l).
\label{eq:gen_win}
\end{equation}

Now we use a small window , $W=2$, to evaluate $\mathbb{E}[n_2]$.
The window of size \( W=2\) spans two syndrome extraction cycle with time stamps as $t_l-\Delta t$ and $t_l-2\Delta t$, where \( p_{ij}(t_l-2\Delta t) = p_1 \) and \( p_{ij}(t_l-\Delta t) = p_2 \) denote the probabilities of observing errors \( e_{ij}(t_l-2\Delta t) = 1 \) and \( e_{ij}(t_l - \Delta t) = 1 \), respectively. Assuming Markovian temporal noise events, \( e_{ij}(t_l- \Delta t) \) and \( e_{ij}(t_l - 2\Delta t) \) are statistically independent. For a window size \( W = 2 \), the expectation value of the number of 1s in the bit string \( B_W \), as given by  Eq.~(\ref{eq:expectation value of n}) , is:
\begin{equation}
\begin{split}
\mathbb{E}[n_2] &= \Big[ 0 \cdot (1 - p_1)(1 - p_2) \\
                &\quad + 1 \cdot \left[ p_1(1 - p_2) + (1 - p_1)p_2 \right] \\
                &\quad + 2 \cdot (p_1 p_2) \Big].
\end{split}
\end{equation}

Here, we enumerate all possible bit strings \( B_W^{(i)} \) for \( W = 2 \), namely \( 00, 01, 10, 11 \), with their respective probabilities:

\begin{equation}
(1 - p_1)(1 - p_2), \quad p_1(1 - p_2), \quad (1 - p_1)p_2, \quad p_1 p_2.
\end{equation}

Simplifying the expectation value yields:

\begin{align}
\mathbb{E}[n_2] &= p_1 - p_1 p_2 + p_2 - p_2 p_1 + 2 p_1 p_2 \nonumber \\
&= p_1 + p_2 \nonumber \\
&= p_{ij}(t_l - \Delta t ) + p_{ij}(t_l - 2\Delta t).
\label{eq:w=2}
\end{align}

Using Eq.~\eqref{eq:window_estimate_edge}, 
 $\langle e_{ij}(t) \rangle$ for $W=2$  is given by $\frac{\mathbb{E}[n_2]}{2}$.
Using the recurrence relation in Eq.~(\ref{eq:gen_win}) and the expectation value of $n$ in Eq.~(\ref{eq:w=2}) for window of size $W=2$, we get the estimated error rate $p_{ij}^{\mathrm{est}}(t_l)$ and  $\langle e_{ij}(t) \rangle$ for window of arbitrary size $W$ as 
\begin{equation}
p_{ij}^{\mathrm{est}}(t_l) = \frac{1}{W} \sum_{k=0}^{W-1} p_{ij}(t_l - W\Delta t + k\Delta t),
\label{eq:append_average}
\end{equation}
which is the temporal average of the ground truth error rates in the window.

\section{Derivation of Standard Deviation using Sliding Window} \label{Appendix B}
We now derive the standard deviation for the sliding window estimator of size $W$. For the random variable $n$ defined in Appendix~\ref{Appendix A}, the variance is given by:
\begin{equation}
    \mathrm{Var}(n) = \mathbb{E}\left[(n - \mathbb{E}[n])^2\right].
\end{equation}
Since the estimated error rate using a window is the expectation value of $n$ normalized by $W$, the variance in the estimated error rate is given by:
\begin{equation}
    \sigma^2 = \frac{\mathbb{E}\left[(n - \mathbb{E}[n])^2\right]}{W^2}.
\end{equation}
This expectation can be expressed as a summation over all possible bit strings $B_W^{(i)}$:
\begin{equation}
    \sigma^2 = \frac{1}{W^2}\sum_{i=0}^{2^W-1} \left(n_{B_W^{(i)}} - \mathbb{E}[n]\right)^2 P(B_W^{(i)}),
\label{eq:sigma2}
\end{equation}
where $n_{B_W^{(i)}}$ is the number of 1s in bit string $B_W^{(i)}$, and $P(B_W^{(i)})$ is its probability as defined in Appendix~\ref{Appendix A}.
Expanding the squared term Eq.~(\ref{eq:sigma2}) yields:

\begin{equation}
    \begin{split}
        \sigma^2 &= \frac{1}{W^2}\sum_{i=0}^{2^W-1} \Big(n_{B_W^{(i)}}^2 P(B_W^{(i)}) + \mathbb{E}[n]^2 P(B_W^{(i)}) 
        \\&~~~~~~~~~~~~~~~
        - 2n_{B_W^{(i)}} \mathbb{E}[n] P(B_W^{(i)}) \Big )
        \\&= \frac{1}{W^2}\left((\sum_{i} n_{B_W^{(i)}}^2 P(B_W^{(i)}) + \mathbb{E}[n]^2 - 2\mathbb{E}[n]^2 \right).
    \end{split}
\label{eq:b3}
\end{equation}
Now we use the relations $\sum_i n_{B_W^{(i)}} P(B_W^{(i)}) = \mathbb{E}[n]$ from Eq.~(\ref{eq:expectation value of n}) and $\sum_i P(B_W^{(i)}) = 1$. The Eq.~\eqref{eq:b3} simplifies to:
\begin{equation}
    \sigma^2 = \frac{1}{W^2}\left(\mathbb{E}[n^2] - \mathbb{E}[n]^2\right).
\end{equation}
Using the mathematical induction approach again, we evaluate $\mathbb{E}[n^2]$ as:
\begin{equation}
\begin{split}
\mathbb{E}[n^2] &= \sum_{k=0}^{W-1} p_{ij}(t_l + (k - W)\Delta t) \\
                &\quad + 2\sum_{k=0}^{W-2} p_{ij}(t_l + (k - W)\Delta t) p_{ij}(t_l + (k - W)\Delta t),
\end{split}
\end{equation}
while $\mathbb{E}[n]^2 = \left(\sum_{k=0}^{W-1} p_{ij}(t_0+(k- W )\Delta t)\right)^2$. The variance therefore becomes:
\begin{equation}
\begin{split}
\sigma^2 &= \frac{1}{W^2}\left[\sum_{k=0}^{W-1} p_{ij}(t_l +(k- W )\Delta t) -  p_{ij}(t_l +(k- W )\Delta t)^2\right] \\
         &= \frac{1}{W^2}\sum_{k=0}^{W-1} p_{ij}(t_l +(k- W )\Delta t)\left(1 - p_{ij}(t_l +(k- W )\Delta t)\right).
\end{split}
\end{equation}
The standard deviation of the windowed estimator is consequently given by:
\begin{equation}
    \sigma_W = \sqrt{\frac{1}{W^2}\sum_{k=0}^{W-1} p_{ij}(t_l +(k- W )\Delta t)\left(1 - p_{ij}(t_l+(k- W )\Delta t)\right)}.
\end{equation}


\clearpage

\bibliography{apssamp}

\end{document}